\documentclass[a4paper,11pt]{article}
\pdfoutput=1 

\usepackage{jheppub} 

\usepackage{epstopdf} 
\usepackage[T1]{fontenc} 
\usepackage[dvipsnames]{xcolor}

\usepackage{graphics}
\usepackage[utf8]{inputenc}
\usepackage{xspace}
\usepackage{amsmath}
\usepackage{amssymb}
\usepackage{url}
\usepackage{rotating}
\usepackage{enumerate}
\usepackage{graphicx}
\usepackage{pdflscape, afterpage, capt-of} 
\usepackage{slashed} 
\usepackage[normalem]{ulem}
\usepackage{nicefrac}
\usepackage[export]{adjustbox} 
\usepackage{makecell} 
\usepackage{subcaption}

\usepackage{def} 

\newcommand{\private}[1]{} 

\title{\boldmath The chirality-flow formalism for the standard model}

\author[a]{Joakim Alnefjord,}
\author[a]{Andrew Lifson,}
\author[a]{Christian Reuschle}
\author[a]{and Malin Sjodahl}

\affiliation[a]{Department of Astronomy and Theoretical Physics, Lund
  University, S{\"o}lvegatan 14A, 223\,62 Lund, Sweden}

\emailAdd{joakim.alnefjord@gmail.com}
\emailAdd{andrew.lifson@thep.lu.se}
\emailAdd{christian.reuschle@thep.lu.se} 
\emailAdd{malin.sjodahl@thep.lu.se}

\abstract{
  In a recent paper we introduced the chirality-flow formalism, a 
  method for
  simple and transparent calculations of Feynman diagrams
  based  on the left- and right-chiral $\mathfrak{sl}(2,\mathbb{C})$ 
  nature of spacetime.
  While our previous work focused on massless QED and QCD at tree-level,
  we here extend the  chirality-flow formalism to the full (tree-level) Standard Model,
  including massive particles and electroweak interactions ---
  for which the $W$-interaction simplifies elegantly due to
  its chiral nature.
  We illustrate how values of Feynman diagrams can be immediately
  written down with some representative examples.
}

\begin{document} 
\preprint{LU-TP 20-51, MCNET-20-25}
\maketitle
\flushbottom

\section{Introduction}
\label{sec:introduction}

In  a recent paper \cite{Lifson:2020pai} 
we introduced the chirality-flow formalism --- a flow-like
method for treating the Lorentz structure of scattering amplitudes --- 
together with its tree-level Feynman rules in massless QED and QCD.
This method builds on the spinor-helicity formalism
\cite{DeCausmaecker:1981jtq,Berends:1981rb,Berends:1981uq,DeCausmaecker:1981wzb,Berends:1983ez,Kleiss:1984dp,Berends:1984gf,Gunion:1985bp,Gunion:1985vca,Kleiss:1985yh,Hagiwara:1985yu,Kleiss:1986ct,Kleiss:1986qc,Xu:1986xb,Gastmans:1987qz,Schwinn:2005pi}
and is inspired by the color-flow decomposition of the color structure
of gluons into fundamental and anti-fundamental representations
\cite{tHooft:1973alw,Kanaki:2000ms,Maltoni:2002mq}.
It similarly decomposes the Lorentz structure of spin-one bosons into
the dotted and undotted left- and right-chiral fields
of the Weyl-van-der-Waerden formalism
\cite{Farrar:1983wk,Berends:1987cv,Berends:1987me,Berends:1988yn,Berends:1988zn,Berends:1989hf,Dittmaier:1993bj,Dittmaier:1998nn,Weinzierl:2005dd},
denoted by dotted and undotted lines respectively, and 
corresponding to the two $\mathfrak{sl}(2,\mathbb{C})$ copies of spacetime.
This allows for recasting Feynman diagrams into simple and intuitive
chirality-flow diagrams, the values of which can be immediately
written down in terms of Lorentz-invariant spinor inner products.

In the present paper we extend the chirality-flow method to the case of
massive particles, focusing on the Standard
Model at tree-level.
This entails dealing with massive spinors, fourvectors, 
and polarization vectors by decomposing them into combinations of massless
spinors \cite{Kleiss:1985yh,Beenakker:1991jk,Dittmaier:1998nn}.
More specifically, we write the Dirac spinor as a linear combination of left- and right-chiral Weyl spinors,
and massive momenta as a linear combination of massless momenta, which are also used to describe the massive polarization vectors.
Using the chirality-flow concepts and notation introduced in
\cite{Lifson:2020pai}, the extension to the full Standard Model then
turns out to be straightforward, and the left-chiral weak part of the
Standard Model simplifies elegantly.
In principle, this extension is to be anticipated, since the ``chirality flows'' 
represent the only Lorentz-invariant quantities at hand, 
the antisymmetric contraction of two left- or right-chiral spinors, i.e.,
the spinor inner products.

This paper is organized as follows: After reviewing some key features of the massless chirality-flow
formalism in \secref{sec:chiralityFlowMassless}, we introduce massive fourvectors,
spinors and polarization vectors in \secref{sec:spinorhelicitymassive}.
The Feynman rules and their consistent application are described
in \secref{sec:chiralityflowrulesmassive},
and values of representative Feynman diagrams are written down in
\secref{sec:examples}. Finally, we conclude in \secref{sec:conclusion}.

\section{Massless chirality-flow}
\label{sec:chiralityFlowMassless}

In this section we go through the basic concepts introduced for massless
chirality-flow in \cite{Lifson:2020pai}. The new structures encountered in
the massive case can be expressed in terms of these massless objects.
Introductions to the spinor-helicity method itself can  be found in
for example \cite{Mangano:1990by,Dixon:1996wi,Dittmaier:1998nn,Weinzierl:2005dd,Weinzierl:2007vk,Dreiner:2008tw,Ellis:2011cr,Peskin:2011in,Elvang:2013cua,Dixon:2013uaa}.

\subsection{Weyl spinors}
\label{sec:masslessspinors}

We begin by recapitulating our notation for massless Weyl spinors. 
After crossing
such that fermions and anti-fermions are \textit{outgoing} 
(i.e.\ fourmomenta point away from the blobs that represent the internal process), 
the four possible types of Weyl spinors are,
\begin{align}
\text{right-chiral fermion}
\;\;,
\quad\;
\lambda_i^{\alpha}  
\;\leftrightarrow\;\, 
\langle i| 
& \;=\; 
\raisebox{-5.5pt}{\includegraphics[scale=0.4]{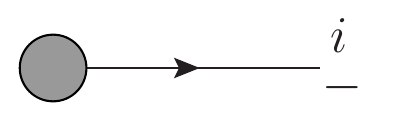}} 
\!\!\!\!=\; 
\raisebox{-5.5pt}{\includegraphics[scale=0.40]{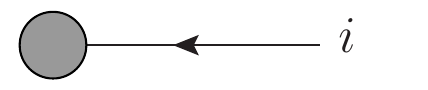}}  
\!\!\!
,
\nonumber
\\
\text{right-chiral anti-fermion} 
\;\;,
\;\;\,
\la_{j,\alpha}  
\;\leftrightarrow\; 
|j\rangle  
& \;=\; 
\raisebox{-5.5pt}{\includegraphics[scale=0.4]{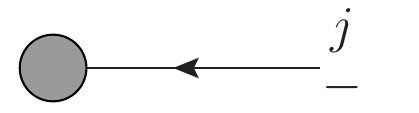}} 
\!\!\!\!=\; 
\raisebox{-5.5pt}{\includegraphics[scale=0.40]{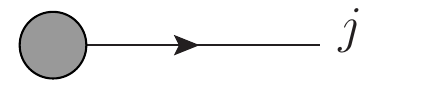}}  
\!\!\!
,
\nonumber
\\
\text{left-chiral fermion} 
\;\;,
\;\;\;
\tilde{\lambda}_{i,\da}  
\;\leftrightarrow\;\,
[i|  
& \;=\; 
\raisebox{-5.5pt}{\includegraphics[scale=0.4]{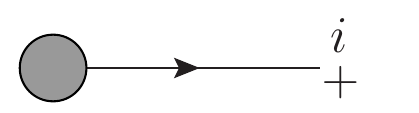}} 
\!\!\!\!=\; 
\raisebox{-5.5pt}{\includegraphics[scale=0.40]{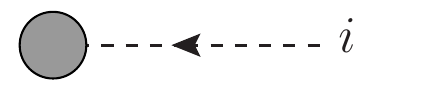}}  
\!\!\!
\nonumber
,
\\
\label{eq:lambdatildelower}
\text{left-chiral anti-fermion} 
\;\;,
\;\;\;\;\,
\tilde{\lambda}_j^{\dot{\alpha}}  
\;\leftrightarrow\; 
|j] 
& \;=\; 
\raisebox{-5.5pt}{\includegraphics[scale=0.4]{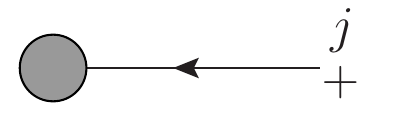}} 
\!\!\!\!=\; 
\raisebox{-5.5pt}{\includegraphics[scale=0.4]{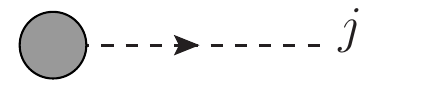}}  
\!\!\!
.
\end{align}
Here, 
the right-chiral Weyl spinors $\langle i|=\langle p_i|$ and $|j\rangle=|p_j\rangle$ 
describe massless fermions and anti-fermions of negative helicity and momentum $p_i$ and $p_j$ respectively, 
while the left-chiral Weyl spinors $[i|=[p_i|$ and $|j]=|p_j]$ describe positive-helicity fermions and anti-fermions.
\footnote{
  If we assume incoming fermions and anti-fermions instead, 
  i.e. fourmomenta 
  pointing towards the blobs that represent the internal process,
  the right-chiral $\langle i|$ and $|j\rangle$ describe massless anti-fermions and fermions of positive helicity and momentum $p_i$ and $p_j$ respectively, 
  while the left-chiral $[i|$ and $|j]$ describe negative-helicity anti-fermions and fermions. 
}

In \eqref{eq:lambdatildelower}
the left graphical rules correspond to the conventional Feynman rules 
(showing the fermion-flow arrows, momentum labels and helicity labels), 
whereas the right graphical rules correspond to the chirality-flow rules 
(showing the chirality-flow arrows, 
and using dotted lines for left-chiral particles with dotted indices 
and solid lines for right-chiral particles with undotted indices).

We recall that spinor indices can be raised and lowered using the Levi-Civita tensor
\begin{equation}
\begin{gathered}
\la_{i,\al} = \epsilon_{\al\be}\la_i^{\be}~, 
\quad 
\tla_{i,\dal} = \epsilon_{\dal\dbe}\tla_i^{\dbe}~,
\quad
\la_i^{\al} = \epsilon^{\al\be}\la_{i,\be}~, 
\quad 
\tla_i^{\dal} = \epsilon^{\dal\dbe}\tla_{i,\dbe}~,
\\
  \epsilon^{12} = -\epsilon^{21} = \epsilon_{21} = -\epsilon_{12} = 1~.
\end{gathered}
\label{eq:epsilon_action}  
\end{equation}
Considering that $\epsilon$ is the only SL(2,$\mathbb{C}$) invariant object,
the definitions for the (antisymmetric,  Lorentz-invariant) spinor inner products follow as\footnote{
  For explicit representations of Weyl spinors and their inner products, 
  see \appref{sec:spinor reps app}.  
}
\begin{alignat}{2}
  \langle i j \rangle 
 &
  \defequal  \la_i^{\al}\la_{j,\al}
  = \epsilon^{\al \be}\la_{i,\be}\la_{j,\al}
  &&= \raisebox{-0.2\height}{\includegraphics[scale=0.4]{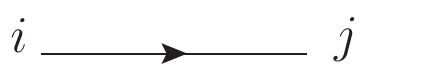}} 
  ~, 
 \nonumber  \\
  [ i j ] 
  &
  \defequal \tla_{i,\da}\tla_j^{\da}
  = \epsilon_{\da \db}\tla_i^{\db}\tla_j^{\da}
  &&= \raisebox{-0.2\height}{\includegraphics[scale=0.4]{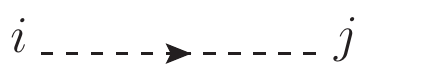}} ~,
    \label{eq:inner products}
\end{alignat}
where $\langle i j\rangle^*=[j i]$ 
for physical momenta (i.e.\ when $p_i^0,p_j^0>0$).
From this,
the chirality-flow representations of a Kronecker delta follow
\begin{align}
  \label{eq:deltas}
  \delta_{\al}^{~\be} = \ \includegraphics[scale=0.40]{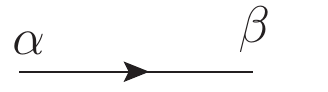}, 
  \qquad 
  \delta_{~\da}^{\db} = \ \includegraphics[scale=0.40]{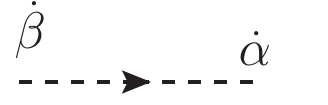}.
\end{align}
We note that the indices in the spinor inner products \eqref{eq:inner products},
as well as in \eqref{eq:deltas} are always intuitively read along the chirality-flow arrows,
leading to the chirality-flow arrows opposing the fermion-flow arrow directions in
\eqref{eq:lambdatildelower}.

For later comparison,
we also write our two-component Weyl spinors
of momentum $p_i$, $p_j$
and helicity $\pm$
in four-component notation
\begin{alignat}{2}
\bar{u}_i^-
=
\bar{v}_i^+
&=
\begin{pmatrix}
0\,\, , &
\raisebox{-5.5pt}{\includegraphics[scale=0.40]{Jaxodraw/ExtSpinorSolidi}} 
\!\!\!\!
\end{pmatrix}~,
\qquad
v_j^-
= u_j^+
&&=
\begin{pmatrix}
0 \\
\raisebox{-5.5pt}{\includegraphics[scale=0.40]{Jaxodraw/ExtSpinorSolidj}}  
\end{pmatrix}~,
\nonumber
\\
\bar{u}_i^+
=
\bar{v}_i^-
&=
\begin{pmatrix}
  \raisebox{-5.5pt}{\includegraphics[scale=0.40]{Jaxodraw/ExtSpinorDottedi}}\!\!,&
  0
\end{pmatrix}~,
\qquad
v_j^+
= u_j^-
&&=
\begin{pmatrix}
\raisebox{-5.5pt}{\includegraphics[scale=0.40]{Jaxodraw/ExtSpinorDottedj}}  
\\0 
\end{pmatrix}~.
\label{eq:fourcomponentweyl}
\end{alignat}

\subsection{Massless fourvectors}
\label{sec:LLfourvectors}
We recall that a fourvector $p$ can be mapped to Hermitian $2\!\times\!2$-matrices,
or bispinors,
\begin{align}
p^{\dal\be}
&\defequal 
p_{\mu}\tau^{\mu,\dal\be} 
=
\frac{1}{\sqrt{2}}
p_{\mu}\si^{\mu,\dal\be} 
\;\;,\quad
\slashed{p}
\defequal   
p_{\mu}\si^{\mu} 
\;\;,
\nonumber
\\ 
\bar{p}_{\al\dbe}
&\defequal 
p_{\mu} \Bar{\tau}^{\mu}_{\al\dbe} 
=
\frac{1}{\sqrt{2}}
p_{\mu} \Bar{\si}^{\mu}_{\al\dbe}
\;\;,\quad \quad\;\;
\bar{\slashed{p}}
\defequal   
p_{\mu}\bar{\si}^{\mu}
\;\;,
\label{eq:p_contr}
\end{align}
where we use a slash notation for the bispinors,
not to be confused with the Feynman slash (see
\appsrefa{sec:diracspinors_app}{sec:weyl spinors app}
for a summary of our notation and conventions and see 
\cite{Lifson:2020pai} for further details).
The index structures of momentum-bispinors are translated to the chirality-flow picture as
\begin{align}
\slashed{p} 
\;\;\leftrightarrow\;\;
\sqrt{2}p^{\da\be} 
\defequal
\raisebox{-0.15\height}{\includegraphics[scale=0.5]{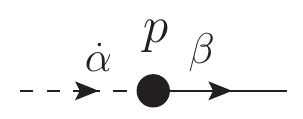}}
\quad\;\;\text{and}\quad\quad
\bar{\slashed{p}} 
\;\;\leftrightarrow\;\;
\sqrt{2}\bar{p}_{\al\db} 
\defequal
\raisebox{-0.15\height}{\includegraphics[scale=0.5]{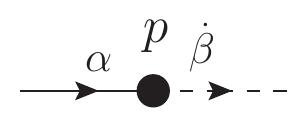}}
\;\;,
\label{eq:p_contr2}
\end{align}
using the momentum-dot notation introduced in \cite{Lifson:2020pai}.
The spinor indices are then contracted with the rest of the amplitude.

If the fourmomentum $p$ is lightlike, its bispinor 
decomposes into an outer product of massless Weyl spinors
\begin{align}
\slashed{p} 
= 
|p]\langle p| 
&\;\;\leftrightarrow\;\;
\sqrt{2}p^{\da\be} 
=
\tla_p^{\da}\la_p^{\be}
\;\;,
\quad\quad\;\text{for}\;\;
p^2=0
\;,\nonumber
\\
\label{eq:Mom two-spinors}
\bar{\slashed{p}} 
= 
|p\rangle [p|
&\;\;\leftrightarrow\;\;
\sqrt{2}\bar{p}_{\al\db} 
= 
\la_{p,\al}\tla_{p,\db}
\;\;,
\quad\text{for}\;\;
p^2=0
\;,
\end{align}
or in the chirality-flow picture, 
\begin{align}
  \label{eq:pdot def}
\raisebox{-0.15\height}{\includegraphics[scale=0.5]{Jaxodraw/FermionPropFlowg}}
\;&=\;\;
\raisebox{-0.20\height}{\includegraphics[scale=0.5]{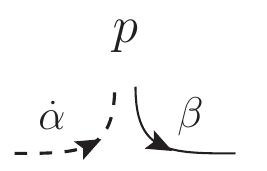}}
,
\quad\text{for}\;\; 
p^2=0
\;,
\nonumber
\\
\raisebox{-0.15\height}{\includegraphics[scale=0.5]{Jaxodraw/FermionPropFlowh}}
\;&=\;\;
\raisebox{-0.20\height}{\includegraphics[scale=0.5]{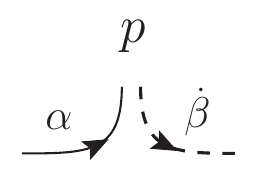}}
,
\quad\text{for}\;\;
p^2=0
\;,
\end{align}
where $p$ in the right graphical rules denotes that the line ends correspond to massless Weyl spinors of momentum $p$ 
(cf. \eqref{eq:lambdatildelower}).

In Feynman diagrams, we often encounter internal fermions for which
the momentum $p$ is a linear combination of lightlike external momenta $p_i$
(in the massless case), 
allowing us to write
\begin{align}
\slashed{p} 
= 
\sum\limits_i
c_i\slashed{p}_i
= 
\sum\limits_i
c_i|p_i]\langle p_i|
&\;\;\leftrightarrow\;\;
\sqrt{2}p^{\da\be} 
=
\sqrt{2}
\sum\limits_i
c_ip^{\da\be}_i
=
\sum\limits_i
c_i\tla_{p_i}^{\da}\la_{p_i}^{\be}
\;,\nonumber
\\
\label{eq:Mom two-spinors_pi}
\bar{\slashed{p}} 
= 
\sum\limits_i
c_i\bar{\slashed{p}}_i
= 
\sum\limits_i
c_i|p_i\rangle [p_i|
&\;\;\leftrightarrow\;\;
\sqrt{2}\bar{p}_{\al\db} 
= 
\sqrt{2}
\sum\limits_i
c_i\bar{p}_{\al\db} 
=
\sum\limits_i
c_i\la_{p_i,\al}\tla_{p_i,\db}
\;,
\end{align}
or in the chirality-flow picture, 
\begin{align}
\raisebox{-0.15\height}{\includegraphics[scale=0.5]{Jaxodraw/FermionPropFlowg}}
\;&=
\;\;\displaystyle\sum\limits_ic_i\raisebox{-0.15\height}{\includegraphics[scale=0.5]{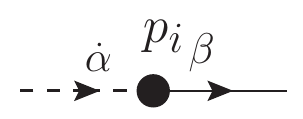}}
\;=\;\;
\displaystyle\sum\limits_ic_i\raisebox{-0.20\height}{\includegraphics[scale=0.5]{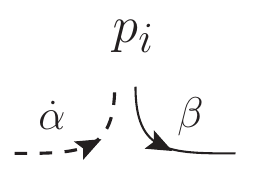}}
,
\nonumber
\\
\raisebox{-0.15\height}{\includegraphics[scale=0.5]{Jaxodraw/FermionPropFlowh}}
\;&=
\;\;\displaystyle\sum\limits_ic_i\raisebox{-0.15\height}{\includegraphics[scale=0.5]{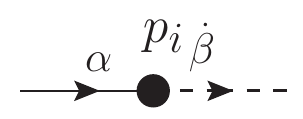}}
\;=\;\; 
\displaystyle\sum\limits_ic_i\raisebox{-0.20\height}{\includegraphics[scale=0.5]{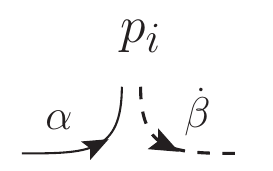}}
.
\label{eq:p_sum}
\end{align}

Also, factors of momentum $p^\mu$ 
which are not contracted with $\tau/\taubar$-matrices  may be encountered.
However, as is argued in section 4.4 of \cite{Lifson:2020pai},
it is always possible to make this contraction within a given Feynman diagram
(see also \secref{sec:arrows and signs}).
Therefore, we are free to write
\begin{align}
p^\mu \rightarrow  p^\mu \tau_\mu^{\da\be}
\leftrightarrow \frac{1}{\sqrt{2}}\slashed{p}
~,
\quad
\mathrm{or}
\quad
p^\mu \rightarrow  p^\mu \taubar_{\mu,\al\db}
\leftrightarrow \frac{1}{\sqrt{2}}\bar{\slashed{p}} 
~,
\end{align}
or in chirality-flow notation
\begin{align}
p^\mu \rightarrow  
\frac{1}{\sqrt{2}}
\raisebox{-0.15\height}{\includegraphics[scale=0.5]{Jaxodraw/FermionPropFlowg}}
~,
\quad
\mathrm{or}
\quad
p^\mu \rightarrow  
\frac{1}{\sqrt{2}}
\raisebox{-0.15\height}{\includegraphics[scale=0.5]{Jaxodraw/FermionPropFlowh}}\;.
\label{eq:pmu chir flow}
\end{align}
Finally,
a massless momentum $p^\mu$ 
can be written in terms of a $\tau/\taubar$ 
matrix and spinors of momentum $p$,
\begin{equation}
p^\mu = \frac{1}{\sqrt{2}}\langle p | \taubar^\mu | p] 
= \frac{1}{\sqrt{2}} [ p | \tau^\mu | p\rangle ~.
\label{eq:Gordon p}
\end{equation}

\subsection{Polarization vectors}
\label{sec:polarization vectors massless}
A key ingredient of the massless chirality-flow formalism is the polarization vector.
It is well-known \cite{Xu:1986xb,Gunion:1985vca}
how to express a massless polarization vector in terms of its momentum $p$
and an arbitrary reference momentum $r$
satisfying $r^2 = 0$ and $r\cdot p \neq 0$
(different choices of $r$ amount to different choices of gauge,
and $r$ may be chosen to simplify a given calculation).
We denote an outgoing polarization vector of helicity $h \in\{+,-\}$
by $\big(\eps^\mu_{h}(p,r)\big)^*$,
while an incoming polarization vector is given by
$\eps^\mu_{h}(p,r)=\big(\eps^\mu_{-h}(p,r)\big)^*$.
For this reason, 
we only consider outgoing polarization vectors and drop the $*$ 
for convenience.

The outgoing polarization vectors are
\begin{alignat}{2}
\eps_+^\mu(p,r) &= \frac{\la_r^{\al}\taubar^\mu_{\al\db}\tla_p^{\db}}{\la_r^{\ga} \la_{p,\ga}}
= \frac{\tla_{p,\da}\tau^{\mu,\da\be}\la_{r,\be}}{\la_r^{\ga} \la_{p,\ga}}
\quad \leftrightarrow \quad
\eps_+^\mu(p,r) &&= \frac{\langle r|\taubar^\mu|p]}{\langle rp\rangle}
= \frac{[p|\tau^\mu|r\rangle}{\langle rp\rangle}~,
    \nonumber \\
    \eps_-^\mu(p,r) &= \frac{\la_p^{\al}\taubar^\mu_{\al\db}\tla_r^{\db}}{\tla_{p,\dga} \tla_{r}^{\dga}}
    = \frac{\tla_{r,\da}\tau^{\mu,\da\be}\la_{p,\be}}{\tla_{p,\dga} \tla_{r}^{\dga}}
    \quad \leftrightarrow \quad
\eps_-^\mu(p,r) &&= \frac{\langle p|\taubar^\mu|r]}{[pr]}
= \frac{[r|\tau^\mu|p\rangle}{[pr]}~,
\label{eq:massless pol vecs mu}
\end{alignat}
which, as for any fourvector,
may be converted into bispinors by contracting with 
$\tau_\mu$ or $\taubar_\mu$
(cf. \eqref{eq:p_contr})
\begin{alignat}{2}
  \eps_{+}^{\dbe\al}(p,r)    
  &=
  \frac{\tla_p^{\db}\la_r^{\al}}{\la_r^{\ga} \la_{p,\ga}}
  \leftrightarrow
  \frac{|p]\langle r|}{\langle rp\rangle}~,    \qquad  
  \bar{\eps}_{+,\be\dal}(p,r)
  &=&  
  \frac{\la_{r,\be}\tla_{p,\da}}{\la_r^{\ga} \la_{p,\ga}}
   \leftrightarrow
  \frac{|r\rangle[p|}{\langle rp\rangle}~,
    \nonumber \\
    \eps_{-}^{\dbe\al}(p,r)
    &=
    \frac{\tla_r^{\db}\la_p^{\al}}{\tla_{p,\dga} \tla_{r}^{\dga}}
    \leftrightarrow
    \frac{|r]\langle p|}{[pr]}~, \qquad
  \bar{\eps}_{-,\be\dal}(p,r)
  &=&
  \frac{\la_{p,\be}\tla_{r,\da}}{\tla_{p,\dga} \tla_{r}^{\dga}}
  \leftrightarrow
  \frac{|p\rangle[r|}{[pr]}~,
    \label{eq:epsilon massless}   
\end{alignat}
from which (using \eqref{eq:lambdatildelower})
the chirality-flow expressions follow
\cite{Lifson:2020pai}
\begin{alignat}{2}
\eps^\mu_{+}(p,r)
&=
\!\raisebox{-0.25\height}{\includegraphics[scale=0.375]{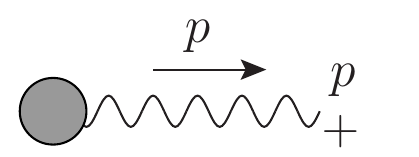}}
\longrightarrow\quad
\frac{1}{\cAngle{rp}}
\!\raisebox{-0.25\height}{\includegraphics[scale=0.375]{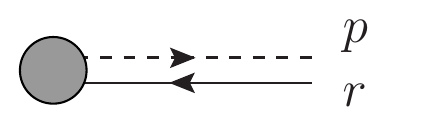}}
\text{or}\quad
&&
\frac{1}{\cAngle{rp}}
\!\raisebox{-0.25\height}{\includegraphics[scale=0.375]{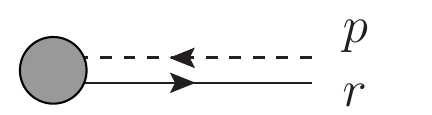}}
\!\!\!\!, \nonumber
\\
\eps^\mu_{-}(p,r)
&=
\!\raisebox{-0.25\height}{\includegraphics[scale=0.375]{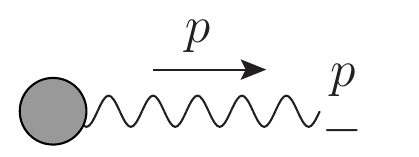}}
\longrightarrow\quad
\frac{1}{\cSquare{p r}}
\raisebox{-0.25\height}{\includegraphics[scale=0.375]{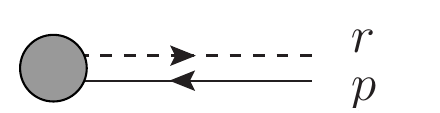}}
\text{or}\quad
&&
\frac{1}{\cSquare{p r}}
\raisebox{-0.25\height}{\includegraphics[scale=0.375]{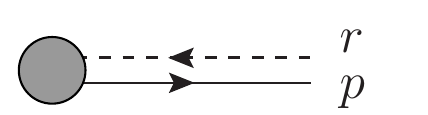}}
\!\!\!\!.
\label{eq:polarisationBispinors+- massless}
\end{alignat}
Note that we have two alternative chirality-flow replacements of
$\eps^\mu_{\pm}(p,r)$. 
In a full chirality-flow diagram we must choose the version which gives a
continuous flow, i.e. which ensures that no arrows point toward or away from
each other along any line in the  chirality-flow diagram. The systematic
treatment of how to achieve this is given in \secref{sec:application}.

\subsection{Linking objects}
\label{sec:remaining massless objects}

The missing ingredients for the massless chirality-flow formalism are
the additional objects which occur in vertices and propagators,
the metric and the $\tau/\taubar$-matrices (from $\gamma$-matrices).

The metric, which always appears contracting two Lorentz indices,
can be translated to a double line, i.e.\ 
parallel dotted and undotted lines with arrows opposing 
\cite{Lifson:2020pai}
\begin{align}
g_{\mu\nu} \leftrightarrow 
\raisebox{-0.25\height}{\includegraphics[scale=0.45]{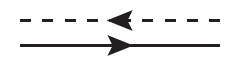}}
\quad \text{or}
\quad \raisebox{-0.25\height}{\includegraphics[scale=0.45]{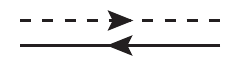}}~,
\label{eq:gmunu sec 2}
\end{align}
where in a given diagram the arrow directions which
give a continuous flow are chosen.
A $\tau/\taubar$-matrix is given by 
\begin{align} \label{eq:tau flow}
\tau^{\mu,\da\be} \leftrightarrow
\includegraphics[scale=0.425,valign=c]{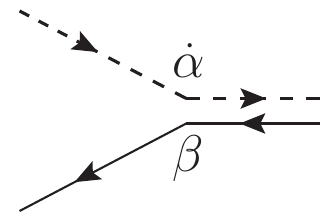}
~,
\qquad
\taubar^\mu_{\al\db} \leftrightarrow
\includegraphics[scale=0.425,valign=c]{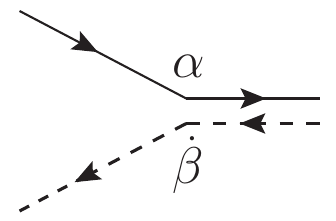}
~,
\end{align}
where the double line to the right in each diagram corresponds to
the $\mu$-index.
At this point, no external spinors are connected to the line ends.

In both cases, the motivation for these replacements lies in the
fact that for a full amplitude, the indices are always
contracted. For the $\tau/\taubar$-matrices the proof also involves
charge conjugation \cite{Lifson:2020pai}.

\section{The chirality-flow formalism with massive particles}
\label{sec:spinorhelicitymassive}

In this section we will set the stage for the massive
chirality-flow formalism, in essence by treating massive
spinors and fourvectors as combinations of
their massless equivalents, and by carefully considering the spin
properties of spin-$1/2$ and spin-1 particles.

\subsection{Spin and helicity}

It is well established that massive and massless particles have rather
different properties under Poincar\'e transformations,
having different little groups
\cite{Wigner:1939cj, Bargmann:1948ck, Weinberg:1995mt}.
Considering only Poincar\'e symmetry, a massless particle can be fully described by its
momentum $p^\mu$ and its helicity $h=\pm J$, where $J$ is the total spin of the particle.
In contrast, 
a massive particle is described by its momentum $p^\mu$, 
total spin $J$, 
and its spin $\Js$ projected onto some axis $\smu$, where
the fourvector $\smu$ must satisfy $\sDotp = 0$ and $\sAxisSq=-1$,
and is often referred to as the polarization vector since $\smu = \eps^\mu_0$
is the longitudinal polarization vector for spin-1 particles
(see e.g.\ \cite{Dittmaier:1998nn,Dreiner:2008tw,Ohlsson:2011zz}).
If the spatial part of $\smu$ points in the direction $\phat=\vec{p}/|\vec{p}|$,  
then $\Js = h$ is the helicity of the particle. 

Analogous to helicity amplitudes for massless particles,
it is possible to define spin amplitudes involving massive particles,
in which each external particle is given a definite spin $\Js$.
However, massive spin amplitudes depend on the directions $\smu$ along which we measure
$\Js$ for each particle, making the choice of $\smu$ physical.
Only the sum of squared spin amplitudes involving a massive particle
are independent of the choice of axis.

In the following sections, we will describe how to separate a massive momentum
into a sum of two massless ones,
and in this process we will also define the spin axis $\smu$. 
We will then describe the massive spinors and polarization vectors
required to calculate massive spin amplitudes
in terms of Weyl spinors of these massless momenta.

We note that this decomposition is not the only way to treat massive particles.
A few years ago an alternate set of massive spinor-helicity 
variables was introduced which made the little group structure explicit \cite{Arkani-Hamed:2017jhn}.
Such variables contain all spin projections of a particle,
and (among other things) have inspired a series of recursion relations 
\cite{Ochirov:2018uyq,Franken:2019wqr,Falkowski:2020aso,Ballav:2020ese},
each with applications to different types of processes. 
While such an approach is related to this work
(see \cite{Ochirov:2018uyq} for how to relate the two approaches),
we here choose to build on the older spinor helicity variables,
and use Feynman diagrams to be as general as possible. 
Nonetheless, adapting the chirality-flow method to the new spinor-helicity
variables is a potentially interesting future research direction.

\subsection{Massive fourvectors}
\label{sec:fourvectors}

A massive fourvector $p$ can be described using two lightlike fourvectors
$p^\flat$ and $q$ satisfying $p^\flat \cdot q\ne 0$.
After choosing $q$ arbitrarily, 
we express the massive fourvector $p$ as
\cite{Kleiss:1985yh,Beenakker:1991jk,Dittmaier:1998nn,Weinzierl:2005dd}
\begin{align}
p
=
p^\flat
+
\alpha q
\;,
\quad
\mathrm{where}
\quad 
\alpha=\frac{p^2}{2p\cdot q}=\frac{p^2}{2p^\flat\cdot q} 
\;\;\xrightarrow{\;\;p^2\rightarrow 0\;\;}\;\;
0
\;.
\label{eq:lightlikevectordecomposition}
\end{align}
Using \eqref{eq:Mom two-spinors_pi}, we have with the above
\begin{align}
\slashed{p} 
&=
|p^\flat]\langle p^\flat|
+
\alpha|q]\langle q|
\;,\nonumber
\\
\label{eq:Mom two-spinors_pflat_q}
\bar{\slashed{p}} 
&=
|p^\flat\rangle[p^\flat|
+
\alpha|q\rangle[q|
\;,
\end{align}
or in graphical notation, 
\begin{align}
\raisebox{-0.15\height}{\includegraphics[scale=0.45]{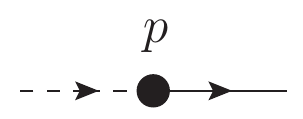}}
\,&=\;\;
\raisebox{-0.20\height}{\includegraphics[scale=0.45]{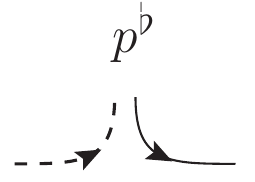}}
\;+\;\;
\alpha\;\raisebox{-0.20\height}{\includegraphics[scale=0.45]{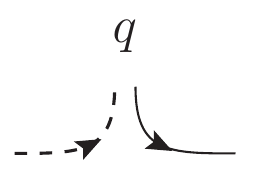}}
,\nonumber
\\
\label{eq:Mom two-spinors_pflat_q_cf}
\raisebox{-0.15\height}{\includegraphics[scale=0.45]{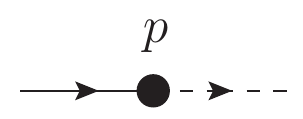}}
\;&=\;\; 
\raisebox{-0.20\height}{\includegraphics[scale=0.45]{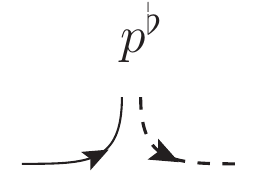}}
\,+\;\;
\alpha\;\raisebox{-0.20\height}{\includegraphics[scale=0.45]{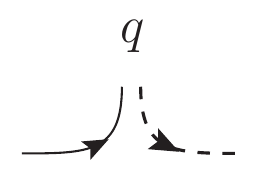}}
,
\end{align}
which trivially reduce to \eqref{eq:pdot def} in the limit
$p^2\rightarrow0\Rightarrow \alpha \rightarrow 0$. 
For fermion propagators in massive Feynman diagrams, 
we may use this to extend \eqref{eq:p_sum} to the case where some of the momenta $p_i$ 
are not lightlike, 
but instead refer to massive on-shell momenta with $p_i^2=m_i^2$.

Using the decomposition into lightlike vectors in \eqref{eq:lightlikevectordecomposition},
we define the axis $\smu$ along which the spin of a
fermion (see \secref{sec:diracspinors})
or vector boson (see \secref{sec:polarizationvectors})
is measured via
\cite{Kleiss:1985yh,Dittmaier:1998nn}
\begin{equation} \label{eq:spinAxis}
\smu = \frac{1}{m}\left( p^{\flat \mu} - \alpha q^\mu \right)
= \frac{1}{m}\left( p^{\mu} - 2\alpha q^\mu \right)~.
\end{equation}
From \eqref{eq:lightlikevectordecomposition}, 
$\smu$ is easily shown to  satisfy the required relations $\sAxisSq=-1$
and $\sDotp = 0$
(see \appref{sec:massive_spin_operator_app}
for more details). 
Therefore, the arbitrary choice of each $q$ is physical,
and different choices of $q$ lead to different spin directions,
implying different spin amplitudes.

\subsubsection{Helicity and the eigenvalue decomposition}
\label{sec:eigenvalue decomposition}

A special case of the decomposition in \eqref{eq:lightlikevectordecomposition} is
the \textit{eigenvalue decomposition}
for which the spatial part of 
$\smu$ points along the direction of motion, $\phat=\vec{p}/|\vp|$.
Therefore, in this decomposition the particles are \textit{helicity} eigenstates.

It is easily shown that both $\slashed{p}$ and $\bar{\slashed{p}}$ have
the (``forward'' and ``backward'') eigenvalues $\eigvals = p^0 \pm|\vec{p}|$,
such that the determinant of $\slashed{p}$ 
is $\eigvalp\eigvalm = m^2$.
Further, the eigenvectors $\eigsSqR{p}, \eigsRan{p}$ 
of $\slashed{p}$ and $\bar{\slashed{p}}$ 
are massless spinors satisfying 
\begin{alignat}{2}
\slashed{p}\eigsSqR{p} &= \eigvals \eigsSqR{p}~,
\qquad \qquad \ 
\bar{\slashed{p}}\eigsRan{p} 
&&= \eigvals \eigsRan{p}~,\nonumber
\\
\cAngle{p_\plus p_\minus} 
&= 
m\sqrt{\frac{\phat^{\perp}}{\phat^{\perp^*}}} 
\defequal m e^{-i\varphi}
\qquad 
\cSquare{p_\minus p_\plus }
&&= 
\cAngle{p_\plus p_\minus}^*
= me^{i\varphi}~, \label{eq:spinor phase eigenvectors}
\end{alignat}
where $\varphi\equiv\varphi_{p_\plus p_\minus}$ is a relative phase (see \appref{sec:useful identities})
and $\phat^{\perp} = \phat^1 +i\phat^2$. 
(If $\phat^{\perp} = 0$, 
$\varphi$ can be set to $0$.)

These spinors can be translated to momenta using 
\eqref{eq:Gordon p},
\begin{align}
\eigsvec{p}
=
\frac{1}{\sqrt{2}} \eigsLan{p}\taubar^\mu\eigsSqR{p}
=
\frac{1}{\sqrt{2}} \eigsSqL{p}\tau^\mu\eigsRan{p}
~, \label{eq:Gordon identity eigenvectors}
\end{align}
with $\vec{p}_{\plus}$ pointing in the same direction as $\phat$
and $\vec{p}_{\minus}$ pointing in the opposite direction,
giving
\begin{align}
\eigvecp{p} = 
\frac{\eigvalp}{2} \big(1,\phat\big)~,
\qquad
\eigvecm{p} = 
\frac{\eigvalm}{2} \big(1,-\phat\big)~.
\end{align}

It is easy to check that a special case of \eqref{eq:lightlikevectordecomposition}
is $p^\mu =p_f^\mu+p_b^\mu$,
corresponding to the choice 
$\alpha = 1$, $q^\mu=\eigvecm{p}$,
and $p^{\flat,\mu}=\eigvecp{p}$.
With this choice, 
\eqref{eq:Mom two-spinors_pflat_q} becomes
\begin{align}
\slashed{p} 
=
\eigSqRp{p}\eigLanp{p}
+
\eigSqRm{p}\eigLanm{p}
\;,
\qquad
\bar{\slashed{p}} 
=
\eigRanp{p}\eigSqLp{p}
+
 \eigRanm{p}\eigSqLm{p}
\;,
\label{eq:Mom two-spinors_eigen_abs}
\end{align}
and the spin fourvector $\smu$ in \eqref{eq:spinAxis} is
\begin{align}
\smu = 
\frac{1}{m}\left(\eigvecp{p} - \eigvecm{p}\right)
=
\frac{1}{m}\left(|\vp| , p^0 \phat \right)~, 
\label{eq:spinAxisHel}
\end{align}
i.e. it is pointing along the direction of motion. 
Therefore, with this decomposition $\Js =h$ measures the helicity.

\subsection{Dirac spinors from massless Weyl spinors}
\label{sec:diracspinors}

Similar to describing massive momenta in terms of momenta $p^\flat$ and $q$, 
we describe massive Dirac spinors in terms of massless Weyl spinors of momenta $p^\flat$ and $q$
\cite{Dittmaier:1998nn,Weinzierl:2005dd,Weinzierl:2007vk,Schwinn:2005pi,Schwinn:2005zm,Schwinn:2007ee,Seth:2016hmv}.
The arbitrary reference fourvector $q$ can be chosen to simplify calculations,
but carries physical meaning as it defines the spin axis $\smu$ in \eqref{eq:spinAxis},
along which the spin is measured.

We let
$u^\Js(p)$,
$\bar{u}^{\Js}(p)$, 
$v^\Js(p)$, 
and  
$\bar{v}^\Js(p)$
denote the momentum space Dirac spinors of 
momentum $p$ 
and spin $\Js=(+,-)\leftrightarrow (\frac{1}{2},-\frac{1}{2}) $
along the $\smu$-axis,
and calculate the spin using the operator
\begin{equation}
  \mathcal{O}_s=
  -\frac{\smu\Si_\mu}{2} =
  \frac{\ga^5\smu\ga_\mu}{2}~, 
\label{eq:Js spinors}
\end{equation}
where $\Si^\mu/2$ is the Lorentz-covariant spin operator for Dirac spinors 
and $\Si\cdot p = 0$  
(see \appref{sec:massive_spin_operator_app} for details).

We may write the Dirac spinors for outgoing fermions $\bar{u}^\Js(p)$
and anti-fermions $v^\Js(p)$ in the chirality-flow picture as\,
\begin{align}
\bar{u}^+(p)
=
\raisebox{-5.5pt}{\includegraphics[scale=0.375]{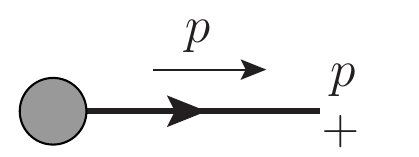}}
\hspace{-2ex}
&=
\bigg(
\!
\raisebox{-5.5pt}{\includegraphics[scale=0.375]{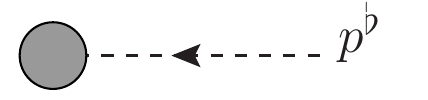}}
\hspace{-2ex}
\;,\;
-e^{i\varphi}
\sqrt{\al}
\raisebox{-5.5pt}{\includegraphics[scale=0.375]{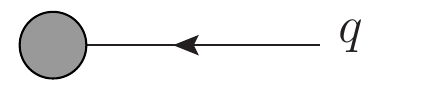}}
\hspace{-2ex}
\bigg)
\;, \nonumber
\\
\bar{u}^-(p)
=
\raisebox{-5.5pt}{\includegraphics[scale=0.375]{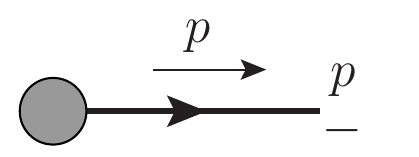}}
\hspace{-2ex}
&=
\bigg(
e^{\!-i\varphi}
\sqrt{\al}
\raisebox{-5.5pt}{\includegraphics[scale=0.375]{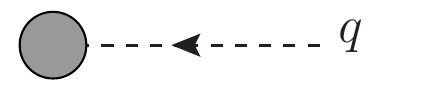}}
\hspace{-2ex}
\;,
\raisebox{-5.5pt}{\includegraphics[scale=0.375]{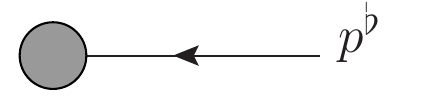}}
\hspace{-1.5ex}
\bigg)
\;, \nonumber
\\
v^+(p)
=
\raisebox{-5.5pt}{\includegraphics[scale=0.375]{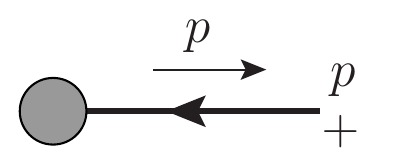}}
\hspace{-2ex}
&=
\begin{pmatrix}
\phantom{\,-e^{i\varphi}\!\sqrt{\alpha}}
\raisebox{-5.5pt}{\includegraphics[scale=0.375]{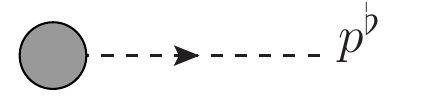}}
\hspace{-1.25ex}
\\
\,-e^{i\varphi}\!\sqrt{\alpha}
\raisebox{-5.5pt}{\includegraphics[scale=0.375]{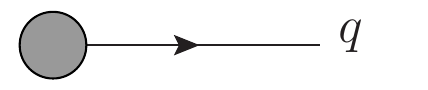}}
\hspace{-1.25ex}
\end{pmatrix}
\;, \nonumber
\\
v^-(p)
=
\raisebox{-5.5pt}{\includegraphics[scale=0.375]{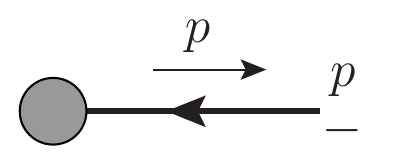}}
\hspace{-2ex}
&=
\begin{pmatrix}
e^{-i\varphi}\!\sqrt{\alpha}
\raisebox{-5.5pt}{\includegraphics[scale=0.375]{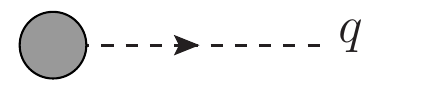}}
\hspace{-1.75ex}
\\
\phantom{e^{-i\varphi}\!\sqrt{\alpha}}
\raisebox{-5.5pt}{\includegraphics[scale=0.375]{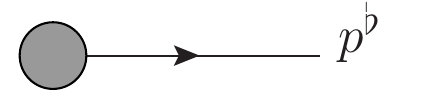}}
\hspace{-1.75ex}
\end{pmatrix}
\;,
\label{eq:ubarvpm_spin_cf}
\end{align}
where in the massless limit $p^\flat=p$ and $\alpha = 0$,
such that we recover \eqref{eq:fourcomponentweyl}.
The factors 
\begin{equation}
e^{i\varphi}\sqrt{\alpha}=\frac{m}{\cAngle{p^\flat q}}\;,
\qquad 
e^{\!-i\varphi}\sqrt{\alpha}=\frac{m}{\cSquare{q p^\flat}}~, \label{eq:spinor phase def with alpha}
\end{equation}
are required such that the spinors satisfy the Dirac equation, and
the phase is given by
\begin{equation}
\cSquare{q p^\flat} =e^{i\varphi}\sqrt{2p^\flat\!\cdot q}\;,
\qquad 
\cAngle{p^\flat q} = e^{-i\varphi}\sqrt{2p^\flat\!\cdot q}~, \label{eq:spinor phase def}
\end{equation}
as in \eqref{eq:spinor phase eigenvectors}.
Explicit forms of the Weyl spinors are given in \eqref{eq:massless Weyl explicit}
(fixing the phase $\varphi$).

Similarly, for incoming anti-fermions $\bar{v}^\Js(p)$ and fermions $u^\Js(p)$ we write
\begin{align}
\bar{v}^+(p)
&=
\raisebox{-5.5pt}{\includegraphics[scale=0.375]{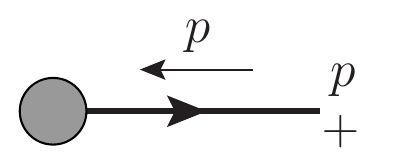}}
\hspace{-2ex}
=
\bigg(
-e^{\!-i\varphi}\!\sqrt{\alpha}
\raisebox{-5.5pt}{\includegraphics[scale=0.375]{./Jaxodraw/CR/ExtSpinorDotted_q}}
\hspace{-2ex}
\;,\;
\raisebox{-5.5pt}{\includegraphics[scale=0.375]{./Jaxodraw/CR/ExtSpinorSolid_pflat}}
\hspace{-1.5ex}
\bigg)
\;, \nonumber
\\
\bar{v}^-(p)
&=
\raisebox{-5.5pt}{\includegraphics[scale=0.375]{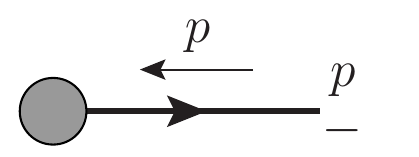}}
\hspace{-2ex}
=
\bigg(
\!\raisebox{-5.5pt}{\includegraphics[scale=0.375]{./Jaxodraw/CR/ExtSpinorDotted_pflat}}
\hspace{-2ex}
\;,\;
e^{i\varphi}\!\sqrt{\alpha}
\raisebox{-5.5pt}{\includegraphics[scale=0.375]{./Jaxodraw/CR/ExtSpinorSolid_q}}
\hspace{-2ex}
\bigg)
\;, \nonumber
\\
u^+(p)
&=
\raisebox{-5.5pt}{\includegraphics[scale=0.375]{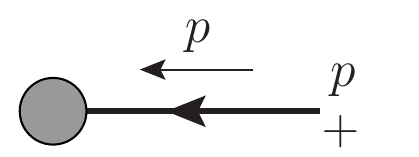}}
\hspace{-2ex}
=
\begin{pmatrix}
-e^{\!-i\varphi}\!\sqrt{\alpha}
\raisebox{-5.5pt}{\includegraphics[scale=0.375]{./Jaxodraw/CR/ExtSpinorAntiDotted_q}}
\hspace{-1.75ex}
\\
\phantom{-e^{\!-i\varphi}\!\sqrt{\alpha}}
\raisebox{-5.5pt}{\includegraphics[scale=0.375]{./Jaxodraw/CR/ExtSpinorAntiSolid_pflat}}
\hspace{-1.75ex}
\end{pmatrix}
\;, \nonumber
\\
u^-(p)
&=
\raisebox{-5.5pt}{\includegraphics[scale=0.375]{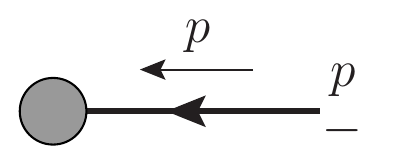}}
\hspace{-2ex}
=
\begin{pmatrix}
\phantom{e^{i\varphi}\!\sqrt{\alpha}}
\raisebox{-5.5pt}{\includegraphics[scale=0.375]{./Jaxodraw/CR/ExtSpinorAntiDotted_pflat}}
\hspace{-1.25ex}
\\
e^{i\varphi}\!\sqrt{\alpha}
\raisebox{-5.5pt}{\includegraphics[scale=0.375]{./Jaxodraw/CR/ExtSpinorAntiSolid_q}}
\hspace{-1.25ex}
\end{pmatrix}
\;.
\label{eq:uvbarpm_spin_cf}
\end{align}
In \eqsrefa{eq:ubarvpm_spin_cf}{eq:uvbarpm_spin_cf} 
the leftmost graphical rules correspond to the conventional Feynman rules 
(showing the fermion-flow arrows, momentum labels and spin labels; the thicker lines imply massive particles), 
and the rightmost graphical rules correspond to the chirality-flow rules (showing the chirality-flow arrow, 
cf. \eqref{eq:lambdatildelower}).

\subsubsection{Helicity eigenstates}
\label{sec:helicity eigenstates}
For our spinors to be helicity eigenstates we use the eigenvalue decomposition detailed in 
\secref{sec:eigenvalue decomposition}, equivalent to choosing
$\alpha=1$,  $q^\mu=\eigvecm{p}$ and $p^{\flat,\mu}=\eigvecp{p}$,
in \eqref{eq:lightlikevectordecomposition}.
Outgoing (anti-)fermions of helicity $\pm$ are then given by
\begin{align}
\bar{u}^+(p)
=
\raisebox{-5.5pt}{\includegraphics[scale=0.375]{Jaxodraw/uBarPl_massive}}
\hspace{-2ex}
&=
\bigg(
\!
\raisebox{-5.5pt}{\includegraphics[scale=0.375]{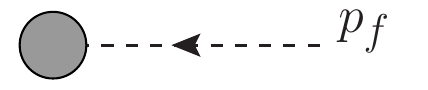}}
\hspace{-2ex}
\;,\;
-e^{i\varphi}
\raisebox{-5.5pt}{\includegraphics[scale=0.375]{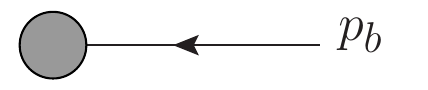}}
\hspace{-2ex}
\bigg)
\;, \nonumber
\\
\bar{u}^-(p)=
\raisebox{-5.5pt}{\includegraphics[scale=0.375]{Jaxodraw/uBarMi_massive}}
\hspace{-2ex}
&=
\bigg(
e^{-i\varphi}
\raisebox{-5.5pt}{\includegraphics[scale=0.375]{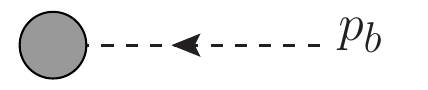}}
\hspace{-2ex}
\;,
\raisebox{-5.5pt}{\includegraphics[scale=0.375]{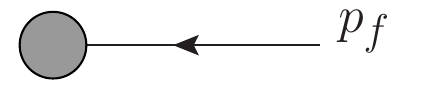}}
\hspace{-1.5ex}
\bigg)
\;, \nonumber
\\
v^+(p)
=
\raisebox{-5.5pt}{\includegraphics[scale=0.375]{Jaxodraw/vPl_massive}}
\hspace{-2ex}
&=
\begin{pmatrix}
\phantom{-e^{i\varphi}}
\raisebox{-5.5pt}{\includegraphics[scale=0.375]{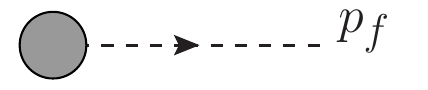}}
\hspace{-1.25ex}
\\
-e^{i\varphi}
\raisebox{-5.5pt}{\includegraphics[scale=0.375]{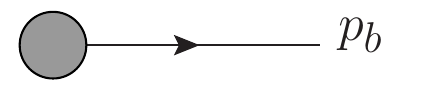}}
\hspace{-1.25ex}
\end{pmatrix}
\;, \nonumber
\\
v^-(p)
=
\raisebox{-5.5pt}{\includegraphics[scale=0.375]{Jaxodraw/vMi_massive}}
\hspace{-2ex}
&=
\begin{pmatrix}
e^{-i\varphi}
\raisebox{-5.5pt}{\includegraphics[scale=0.375]{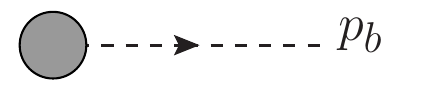}}
\hspace{-1.75ex}
\\
\phantom{e^{-i\varphi}}
\raisebox{-5.5pt}{\includegraphics[scale=0.375]{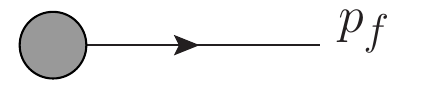}}
\hspace{-1.75ex}
\end{pmatrix}
\;,
\label{eq:vubarpm_hel_cf}
\end{align}
while incoming spinors are given by
\begin{align}
\bar{v}^+(p)
=
\raisebox{-5.5pt}{\includegraphics[scale=0.375]{Jaxodraw/vBarPl_massive}}
\hspace{-2ex}
&=
\bigg(
-e^{-i\varphi}
\raisebox{-5.5pt}{\includegraphics[scale=0.375]{./Jaxodraw/CR/ExtSpinorDotted_pb}}
\hspace{-2ex}
\;,\;
\raisebox{-5.5pt}{\includegraphics[scale=0.375]{./Jaxodraw/CR/ExtSpinorSolid_pf}}
\hspace{-1.5ex}
\bigg)
\;, \nonumber
\\
\bar{v}^-(p)
=
\raisebox{-5.5pt}{\includegraphics[scale=0.375]{Jaxodraw/vBarMi_massive}}
\hspace{-2ex}
&=
\bigg(
\!\raisebox{-5.5pt}{\includegraphics[scale=0.375]{./Jaxodraw/CR/ExtSpinorDotted_pf}}
\hspace{-2ex}
\;,\;
e^{i\varphi}
\raisebox{-5.5pt}{\includegraphics[scale=0.375]{./Jaxodraw/CR/ExtSpinorSolid_pb}}
\hspace{-2ex}
\bigg)
\;, \nonumber
\\
u^+(p)
=
\raisebox{-5.5pt}{\includegraphics[scale=0.375]{Jaxodraw/uPl_massive}}
\hspace{-2ex}
&=
\begin{pmatrix}
-e^{-i\varphi}
\raisebox{-5.5pt}{\includegraphics[scale=0.375]{./Jaxodraw/CR/ExtSpinorAntiDotted_pb}}
\hspace{-1.75ex}
\\
\phantom{-e^{-i\varphi}}
\raisebox{-5.5pt}{\includegraphics[scale=0.375]{./Jaxodraw/CR/ExtSpinorAntiSolid_pf}}
\hspace{-1.75ex}
\end{pmatrix}
\;, \nonumber
\\
u^-(p)
=
\raisebox{-5.5pt}{\includegraphics[scale=0.375]{Jaxodraw/uMi_massive}}
\hspace{-2ex}
&=
\begin{pmatrix}
\phantom{e^{i\varphi}}
\raisebox{-5.5pt}{\includegraphics[scale=0.375]{./Jaxodraw/CR/ExtSpinorAntiDotted_pf}}
\hspace{-1.25ex}
\\
e^{i\varphi}
\raisebox{-5.5pt}{\includegraphics[scale=0.375]{./Jaxodraw/CR/ExtSpinorAntiSolid_pb}}
\hspace{-1.25ex}
\end{pmatrix}
\;,
\label{eq:uvbarpm_hel_cf}
\end{align}
where we use \eqref{eq:spinor phase eigenvectors} 
to define both the Weyl spinors (for which explicit forms are given in \eqref{eq:pfb explicit})
and the phase $\varphi$,
\begin{equation}
e^{i\varphi}=\frac{m}{\cAngle{p_\plus p_\minus}}~,
\qquad 
e^{\!-i\varphi}=\frac{m}{\cSquare{p_\minus p_\plus}}~. \label{eq:spinor phase def with alpha hel}
\end{equation}

\subsection{Polarization vectors}
\label{sec:polarizationvectors}

It is also well-known how to use \eqref{eq:lightlikevectordecomposition}
to describe massive polarization vectors
\cite{Xu:1986xb, Dittmaier:1998nn,Weinzierl:2005dd,Weinzierl:2007vk,Schwinn:2005pi,Schwinn:2005zm,Schwinn:2007ee,Seth:2016hmv}.
Let $\big(\eps^\mu_{s}(p)\big)^*$ 
denote the polarization vectors of outgoing vector bosons 
with momentum $p$ 
and spin (polarization) label $\Js \in\{+,0,-\}$ along the
spin axis $\smu$.
Since incoming polarization vectors are described by
$\eps^\mu_{s}(p)=\big(\eps^\mu_{-s}(p)\big)^*$,
we will again only consider outgoing polarization vectors,
and drop the $*$ for convenience.

The positive and negative polarization vectors may be written
in a form analogous to the massless case, \eqref{eq:epsilon massless},
\begin{alignat}{2}
  \eps_{+}^{\dbe\al}(p^\flat,q)    
  &=
  \frac{\tla_{p^\flat}^{\db}\la_q^{\al}}{\la_q^{\ga} \la_{p^\flat,\ga}}
  \leftrightarrow
  \frac{|p^\flat]\langle q|}{\langle qp^\flat\rangle}~,    \qquad  
  \bar{\eps}_{+,\be\dal}(p^\flat,q)
  &=&  
  \frac{\la_{q,\be}\tla_{p^\flat,\da}}{\la_q^{\ga} \la_{p^\flat,\ga}}
   \leftrightarrow
  \frac{|q\rangle[p^\flat|}{\langle q p^\flat\rangle}~,
    \nonumber \\
    \eps_{-}^{\dbe\al}(p^\flat,q)
    &=
    \frac{\tla_q^{\db}\la_{p^\flat}^{\al}}{\tla_{p^\flat,\dga} \tla_{q}^{\dga}}
    \leftrightarrow
    \frac{|q]\langle p^\flat|}{[p^\flat q]}~, \qquad
  \bar{\eps}_{-,\be\dal}(p^\flat,q)
  &=&
  \frac{\la_{p^\flat,\be}\tla_{q,\da}}{\tla_{p^\flat,\dga} \tla_{q}^{\dga}}
  \leftrightarrow
  \frac{|p^\flat\rangle[q|}{[p^\flat q]}~,
    \label{eq:pm fourpolarizationMassive}   
\end{alignat}
where, unlike the reference spinor $r$ in \eqref{eq:epsilon massless},
the arbitrary reference spinor $q$ carries physical meaning 
as it defines the spin axis $s^\mu$.
These expressions can be straightforwardly translated to the graphical representation 
\begin{alignat}{2}
\eps^\mu_{+}(p^\flat,q)
&=
\!\raisebox{-0.25\height}{\includegraphics[scale=0.375]{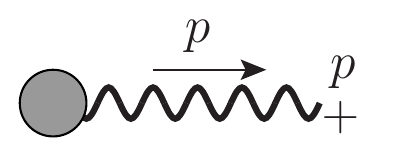}}
\longrightarrow\quad
\frac{1}{\cAngle{qp^\flat}}
\!\raisebox{-0.25\height}{\includegraphics[scale=0.375]{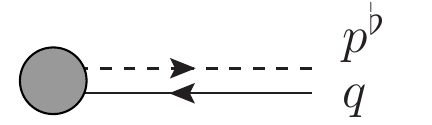}}
\text{or}
&&
\frac{1}{\cAngle{qp^\flat}}
\!\raisebox{-0.25\height}{\includegraphics[scale=0.375]{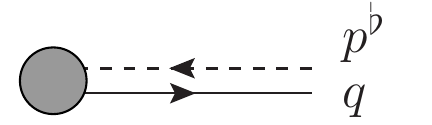}}
\!\!\!\!, \nonumber
\\
\eps^\mu_{-}(p^\flat,q)
&=
\!\raisebox{-0.25\height}{\includegraphics[scale=0.375]{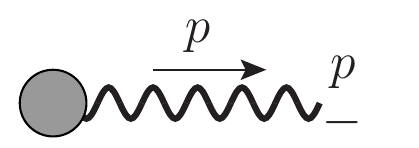}}
\longrightarrow\quad
\frac{1}{\cSquare{p^\flat q}}
\raisebox{-0.25\height}{\includegraphics[scale=0.375]{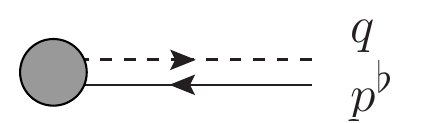}}
\text{or}\quad
&&
\frac{1}{\cSquare{p^\flat q}}
\raisebox{-0.25\height}{\includegraphics[scale=0.375]{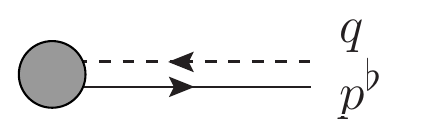}}
\!\!\!\!,
\label{eq:polarisationBispinors+-}
\end{alignat}
where we use thicker wavy lines to imply that the vector bosons are massive,
and the arrow directions which give a continuous flow in a given diagram are chosen.

The additional, longitudinal, polarization vector equals
the spin vector
\begin{eqnarray}
\label{eq:eps0}
\eps^\mu_{0}(p^\flat,q)
&=&
\smu
=
\frac{1}{m}\left( p^{\flat \mu} - \alpha q^\mu \right)
\leftrightarrow
\frac{1}{m\sqrt{2}}
\Big(|p^\flat]\langle p^\flat | - \alpha |q]\langle q|\Big)
\ \
\mathrm{ or }
\ \
\frac{1}{m\sqrt{2}}
\Big(|p^\flat\rangle [p^\flat|-\alpha |q\rangle [q|\Big)
~,
\end{eqnarray}
and using the momentum-dot notation, we can translate this into
\begin{equation}
\eps^\mu_{0}(p^\flat,q)
=
\!\raisebox{-0.25\height}{\includegraphics[scale=0.375]{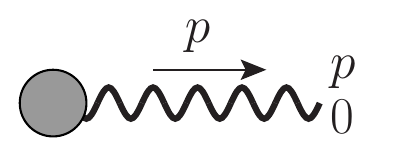}}
\longrightarrow\quad
\frac{1}{m\sqrt{2}}
\!\raisebox{-0.25\height}{\includegraphics[scale=0.375]{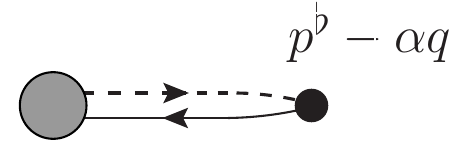}}
\text{or}\quad\;\,
\frac{1}{m\sqrt{2}}
\!\raisebox{-0.25\height}{\includegraphics[scale=0.375]{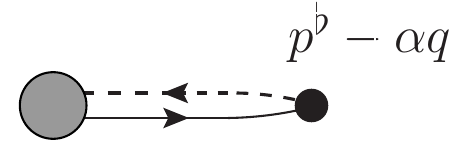}}
,
\label{eq:eps0 dot}
\end{equation}
where in this case there is no massless equivalent.
These graphical representations come from using the first form of
$\eps^\mu_{0}$ in \eqref{eq:eps0}.
Alternatively, we could have
rewritten $\eps^\mu_{0}$ as a linear combination of two spinors as in
\eqref{eq:polarisationBispinors+-} using one of the last two expressions
instead.

\subsubsection{Helicity eigenstates}
To describe external vector bosons as helicity eigenstates we choose 
$\alpha=1$,  $q^\mu=\eigvecm{p}$
and $p^{\flat,\mu}=\eigvecp{p}$, such that $p=p_\plus+ p_\minus$,
as in \secref{sec:eigenvalue decomposition}.
The chirality-flow representations for these states are then
\begin{alignat}{5}
\eps^\mu_{+}(p)=\eps^\mu_{+}(p_\plus, p_\minus)
&=
\!\raisebox{-0.25\height}{\includegraphics[scale=0.375]{Jaxodraw/ZExtPl}}
&&\longrightarrow\quad
&& 
\frac{1}{\cAngle{p_\minus p_\plus}}
\!\raisebox{-0.25\height}{\includegraphics[scale=0.375]{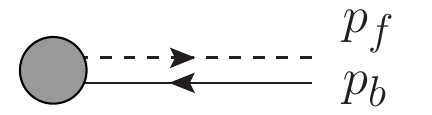}}
&&\text{or}\quad
&&
\frac{1}{ \cAngle{p_\minus p_\plus}}
\!\raisebox{-0.25\height}{\includegraphics[scale=0.375]{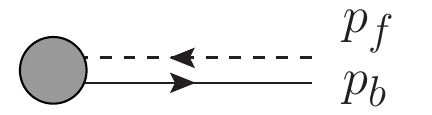}}
, \nonumber
\\
\eps^\mu_{-}(p)=\eps^\mu_{-}(p_\plus, p_\minus)
&=
\!\raisebox{-0.25\height}{\includegraphics[scale=0.375]{Jaxodraw/ZExtMi}}
&&\longrightarrow\quad
&&
\frac{1}{\cSquare{p_\plus p_\minus}}
\raisebox{-0.25\height}{\includegraphics[scale=0.375]{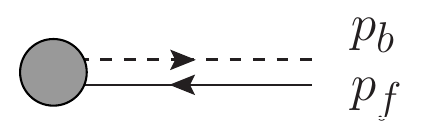}}
&&\text{or} \,\,\,\,\,\, 
&&
\frac{1}{\cSquare{p_\plus p_\minus}}
\raisebox{-0.25\height}{\includegraphics[scale=0.375]{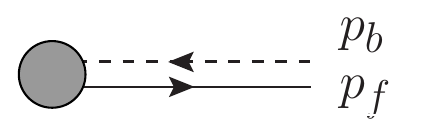}}
, \nonumber
\\
\eps^\mu_{0}(p)=\eps^\mu_{0}(p_\plus, p_\minus)
&=
\!\raisebox{-0.25\height}{\includegraphics[scale=0.375]{Jaxodraw/ZExt0}}
&&\longrightarrow\quad
&&
\frac{1}{m\sqrt{2}}
\, \raisebox{-0.25\height}{\includegraphics[scale=0.375]{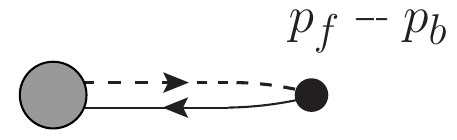}}
\!
&&\text{or}\quad
&&
\frac{1}{m\sqrt{2}}
\!\raisebox{-0.25\height}{\includegraphics[scale=0.375]{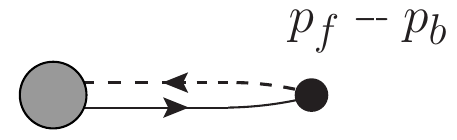}}
,
 \label{eq:polarisationBispinorsHelicity}
\end{alignat}
where in the massless case the longitudinal polarization disappears,
 $p_\plus \rightarrow p$ and the reference momentum
$q = p_\minus$ becomes an \textit{unphysical} gauge choice
(see \cite{Lifson:2020pai} for more details).

\section{Chirality-flow Feynman rules with massive particles}
\label{sec:chiralityflowrulesmassive}

In this section, we describe the (tree-level) electroweak vertices and propagators. 
For convenience,
all tree-level Standard Model rules are also collected in
\appref{sec:rosettastones_app}, 
while a derivation of the QCD chirality-flow rules can be found in 
\cite{Lifson:2020pai}.

We treat the momenta of all particles as outgoing, and use 't Hooft-Feynman gauge.
Further details about the electroweak sector can be found\footnote{
We note though that our conventions differ from those in \cite{Denner:1991kt}.
Following the notation of \cite{Romao:2012pq},
we adopt the conventions of \cite{Peskin:1995ev} 
and use $\eta_e = -1$ in photon vertices, 
while \cite{Denner:1991kt} uses $\eta_e = 1$
(we also differ from \cite{Denner:1991kt} in the definition of $\eta'$
and $\eta_\theta$ of \cite{Romao:2012pq}).
} 
in
\cite{Denner:1991kt}.
We also remark that --- while we focus on tree-level --- similar chirality-flow
structures of vertices and propagators are also expected to be applicable
for loop methods 
which rely on four-dimensional tree objects. 
For physics beyond the Standard Model, 
we note that it is possible to use chirality-flow to describe 
any theory
for which the Lorentz structure can be written down in terms of momenta,
the (Minkowski) metric, Dirac (Pauli) matrices, massless Weyl spinors,
and polarization vectors. 
For this reason, the extension to for example 
$R_\xi$ or axial gauges should also be straightforward.

\subsection{Vertices}
\label{sec:vertices}

\subsubsection{Triple vertices}
The Lorentz structure of the fermion-vector vertices is in principle
unchanged compared to the massless case,
and separates nicely into two chiral parts.
Using \eqref{eq:tau flow} to describe the $\tau/\taubar$ matrices,
we have
\begin{align}
  \includegraphics[scale=0.45,valign=c]{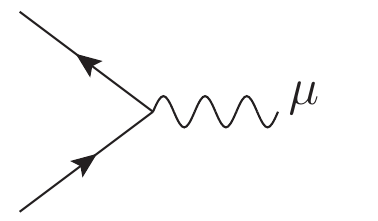} &=
  ie \gamma^\mu (C_L P_L+C_RP_R)  
 \label{eq:fermion_photon_vertex}  
  \\
  &\leftrightarrow
  ie \begin{pmatrix}
    0&   C_{R}\sqrt{2}\tau^{\mu,\da\be}
    \\
     C_{L}\sqrt{2}\taubar^{\mu}_{\al\db} & 0
  \end{pmatrix}
  \leftrightarrow
 ie \sqrt{2}\begin{pmatrix}
   0&   C_{R} \raisebox{-0.35\height}{\includegraphics[scale=0.425]{./Jaxodraw/TauVertexHighb}}
   \\
   C_{L} \raisebox{-0.4\height}{\includegraphics[scale=0.425]{./Jaxodraw/TauVertexLowa}} & 0
 \end{pmatrix}~,\nonumber
\end{align}
with, in general, different couplings to the left and right chiral parts, $C_{L/R}$.
In particular, $C_R=0$ for $W$-interactions, corresponding
to the fact that $W$-bosons only couple to particles in
the left chiral gauge group.\footnote{
 Taking the word chirality to refer to the left-
 and right-chiral representations of the Lorentz group,
 and referring to the spinors
 $[i|,|i]$ as left-chiral
 and
 $\langle j|, |j\rangle$ as right-chiral (as is often done in the literature),
 there is a mismatch w.r.t. how the word chiral is
 used for gauge groups, since clearly the off-diagonal
 nature of the interaction in \eqref{eq:fermion_photon_vertex}
 couples the spinors $\langle j|$ and $|i]$ (in proportion to $C_L$)
and the spinors $[i|$ and $|j\rangle$ (in proportion to $C_R$).
} For the weak SU$(2)_L$ sector, Feynman rules are thus
further simplified in the chirality-flow formalism.

Assuming a diagonal flavor matrix,
$C_L=1/(\sqrt{2}\sin \theta_W)$
for left chiral leptons coupling to $W$, whereas the coupling to
the $Z$-boson is given by
$-Q_f \sin \theta_W/\cos \theta_W $
for right chiral fermions, and by 
$(T_3^f-Q_f \sin^2 \theta_W)/(\cos\theta_W \sin\theta_W)$
for left chiral fermions, 
where $T_3^f$ is the eigenvalue of the third weak isospin generator and
$\theta_W$ is the Weinberg angle. 
The coupling to the photon is given by $C_L=C_R=Q_F$.

Fermions may also couple to scalars,
giving a fermion-scalar vertex of the form
\begin{align}
  \raisebox{-0.4\height}{\includegraphics[scale=0.35]{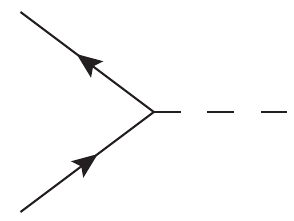}}
   &=ie(C_L P_L + C_R P_R) \nonumber\\
   &\leftrightarrow ie \begin{pmatrix}
   C_L \delta_{~\da}^\db & 0 \\
   0 & C_R \de_\al^{~\be}
\end{pmatrix}    
\leftrightarrow ie \begin{pmatrix}
C_L \raisebox{-0.4\height}{\includegraphics[scale=0.35]{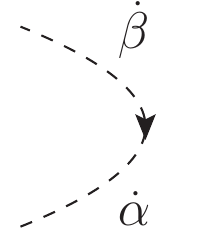}} & 0 \\
 0 &  C_R \raisebox{-0.4\height}{\includegraphics[scale=0.35]{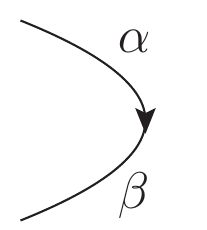}} 
 \end{pmatrix}~, \label{eq:Sff}
\end{align}
where we used \eqref{eq:deltas}
to describe the Kronecker deltas, 
and where,
for a general scalar coupling, the constants $C_L$ and $C_R$
may be different.
In the Standard Model the only known scalar, the Higgs, couples with
$C_L=C_R=-m_f/(2 \sin\theta_W m_W)$
in terms of the fermion mass $m_f$ and the $W$ mass $m_W$.

Here we note that the appearance of a single dotted or undotted line in 
\eqref{eq:Sff}
may give rise to an odd number of chirality-flow lines in a graph,
meaning that the sign flips from reversing
chirality-flow arrows have to be accounted for (as described below in
\secref{sec:arrows and signs}).
We also remark that although it may appear in the graphical representation
to the right as if the scalar has no effect at all, its momentum
enters via momentum conservation.

A Lorentz scalar may also couple to two vector bosons in a vertex of
the form
\begin{equation}
  \label{eq:VVS}
  \raisebox{-0.4\height}{\includegraphics[scale=0.35]{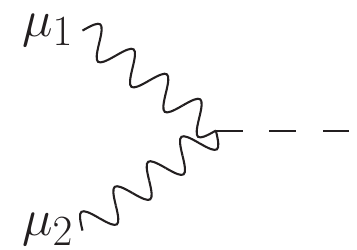}}\;
  =
  ieC_{VVS} g_{\mu_1 \mu_2}
  \leftrightarrow\;
  ieC_{VVS} \raisebox{-0.4\height}{\includegraphics[scale=0.35]{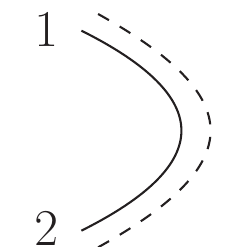}}\;,
\end{equation}
where for brevity we write the metric without arrows,
to indicate that either choice in \eqref{eq:gmunu sec 2} 
may be required 
(recall that flow arrows must never oppose or point away from each other).
Here, again, the presence of the scalar 
manifests itself only via momentum conservation.
In the tree-level Standard Model, this Lorentz structure is applicable
to the Higgs coupling to $W$ and $Z$, for which
$C_{WWh} = m_W/\sin(\theta_W)$ and
$C_{ZZh} = m_Z/(\sin\theta_W\cos\theta_W) 
= m_W/(\sin\theta_W\cos^2\theta_W)$.

There is also a pure scalar vertex,
\begin{equation}
  \includegraphics[scale=0.35,valign=c]{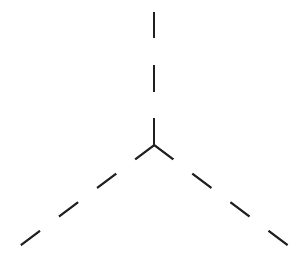}
  \!\! =
  ieC_{SSS}\;,
\end{equation}
with completely trivial Lorentz structure,
and $C_{hhh}=-3 m_h^2/(2 \sin \theta_W m_W)$.
For the non-Abelian vertex with three spin-1 bosons, the Lorentz
structure is as for QCD,
\begin{align}
  \includegraphics[scale=0.35,valign=c]{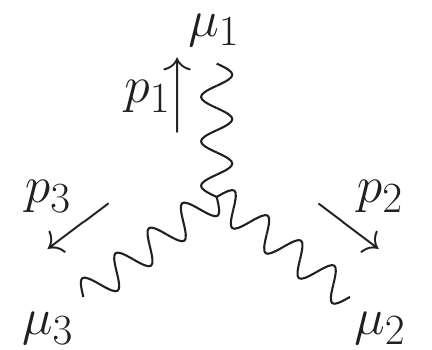}
  &=ieC_{V_1V_2V_3} \left[ g_{\mu_1 \mu_2} (p_1 - p_2)_{\mu_3} + g_{\mu_2 \mu_3} (p_2 - p_3)_{\mu_1} + g_{\mu_3 \mu_1} (p_3 - p_1)_{\mu_2} \right] \nonumber \\
  &\leftrightarrow ieC_{V_1V_2V_3} \frac{1}{\sqrt{2}} \left( 
  \includegraphics[scale=0.35,valign=c]{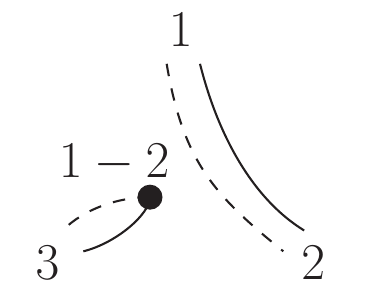}+
  \includegraphics[scale=0.35,valign=c]{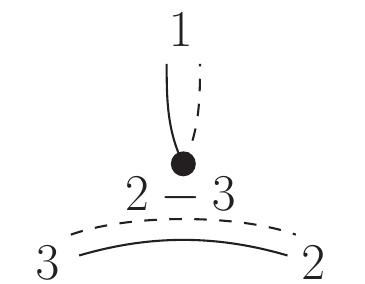}+
  \includegraphics[scale=0.35,valign=c]{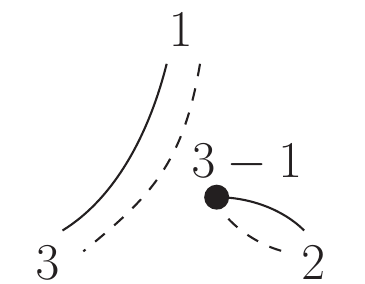} 
  \right)\;,
  \label{eq:VVV}
\end{align}
where \eqref{eq:pmu chir flow} has been used for the momenta $p_1-p_2$ etc.,
and where $C_{\gamma W^+W^-}=-1$ and $C_{ZW^+W^-}=-\cos(\theta_W)/\sin(\theta_W)$.

To complete the list of trivalent vertices, we give the Lorentz
structure of two scalars coupling to a spin-1 particle
\begin{equation}
  \raisebox{-0.4\height}{\includegraphics[scale=0.35]{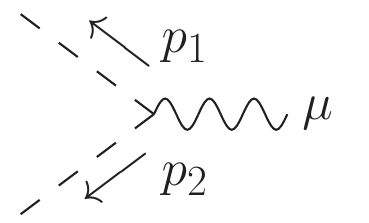}} 
  \!\! =
  ieC_{SSV}(p_1 - p_2)_\mu \leftrightarrow ieC_{SSV} \frac{1}{\sqrt{2}}
  \raisebox{-0.4\height}{\includegraphics[scale=0.35]{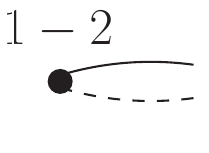}}\;.
\end{equation}
This vertex enters in the Standard Model only at loop level, but
we include
it here to complete the list of Standard Model Lorentz structures
(note that the ghost-ghost-vector vertex has the same chirality-flow
structure but a different momentum-dot argument).

\subsubsection{Four-boson vertices}

We start with the vertex with four vector bosons,
which has a Lorentz structure similar to the color-flow version of the four-gluon
vertex (see section 5.1 in \cite{Lifson:2020pai}),
\begin{eqnarray}
  \includegraphics[scale=0.35,valign=c]{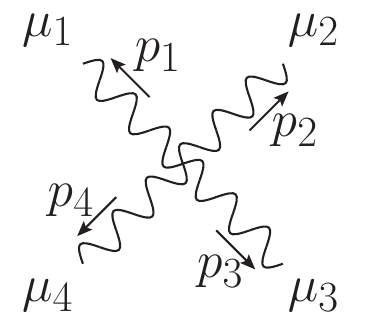}
  &=& ie^2C_{V_1V_2V_3V_4} \left( 2g_{\mu_1 \mu_3} g_{\mu_4 \mu_2} - g_{\mu_1 \mu_2} g_{\mu_3 \mu_4} - g_{\mu_1 \mu_4} g_{\mu_2 \mu_3} \right)\nonumber \\
  &\leftrightarrow& ie^2C_{V_1V_2V_3V_4} \left( 2\includegraphics[scale=0.35,valign=c]{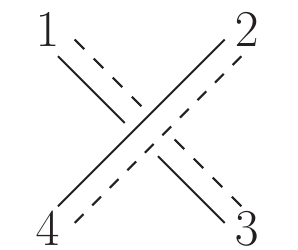}-\includegraphics[scale=0.35,valign=c]{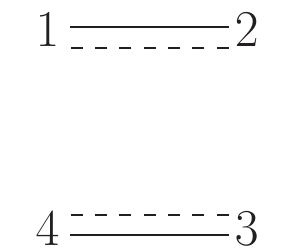}-\includegraphics[scale=0.35,valign=c]{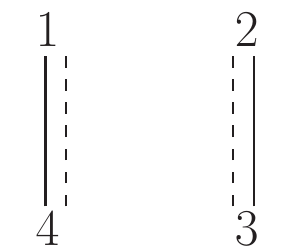} \right)\;.
\end{eqnarray}
Here, the value of $C_{V_1 V_2 V_3 V_4}$ depends on the involved electroweak bosons.
Specifically, we have
$C_{W^+ W^- W^+ W^-}=1/\sin^2\theta_W$,
$C_{W^+ Z W^- Z }=-\cos^2\theta_W/\sin^2\theta_W$,
$C_{W^+ Z W^- \gamma }=-\cos\theta_W/\sin\theta_W$ and
$C_{W^+ \gamma W^- \gamma }=-1$.

We can also have a vertex with two scalars and two vectors,
connected with the chirality flow
\begin{equation}
  \includegraphics[scale=0.35,valign=c]{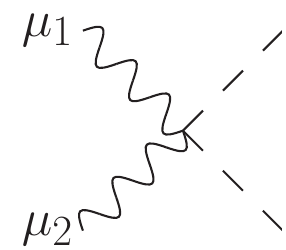}
  =ie^2C_{VVSS} \,g_{\mu_1 \mu_2}
  \leftrightarrow ie^2C_{VVSS} \includegraphics[scale=0.35,valign=c]{Jaxodraw/vertChirCurvedBoson}\;,
\end{equation}
where in the Standard Model
$C_{WWhh}=1/(2 \sin^2\theta_W$) and
$C_{ZZhh}=1/(2 \sin^2\theta_W \cos^2\theta_W)$.

Finally, we have the quartic scalar vertex, with trivial Lorentz structure
\begin{equation}
  \includegraphics[scale=0.35,valign=c]{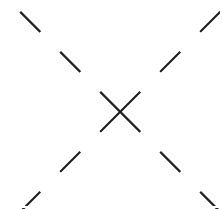} = ie^2C_{SSSS}\;.
\end{equation}
In the tree-level Standard Model, this vertex is applicable to the
quartic Higgs coupling with $C_{hhhh}=-3m_h^2/(4 \sin^2 \theta_W m_W^2)$.

\subsection{Propagators}
\label{sec:propagators}
Adding masses to the chirality-flow picture gives rise to increasing complexity
for the Fermion propagator, 
\begin{equation}
    \raisebox{-0.1\height}{\includegraphics[scale=0.4]{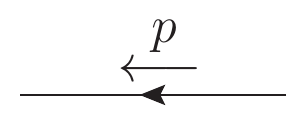}} 
   =
   \frac{i(p_\mu\gamma^\mu+m)}{p^2-m^2}
  \leftrightarrow
  \frac{i}{p^2-m^2} 
  \begin{pmatrix} 
    m {\delta^{\da}}_{\db} & \sqrt{2}p^{\da \be} \\ 
    \sqrt{2}\bar{p}_{\al \db} &  m {\delta_{\al}}^{\be}
  \end{pmatrix}\;,
\end{equation}
which can be drawn in terms of a combination of four different
chirality flows, coupling the left- and right-chiral components,
\begin{equation}
  \label{eq:fermion prop pic}
    \raisebox{-0.1\height}{\includegraphics[scale=0.4]{Jaxodraw/FermionProp_Feyn}} 
    \leftrightarrow \frac{i}{p^2-m^2} 
  \begin{pmatrix} 
    m {\delta^{\da}}_{\db} & \sqrt{2}p^{\da \be} \\ 
    \sqrt{2}\bar{p}_{\al \db} &  m {\delta_{\al}}^{\be}
  \end{pmatrix}\;
  \leftrightarrow 
  \frac{i}{p^2-m^2} 
  \begin{pmatrix} 
    m\includegraphics[scale=0.4]{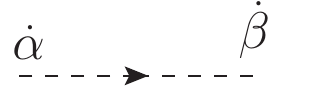} & \includegraphics[scale=0.4]{./Jaxodraw/FermionPropFlowg} \\ 
    \includegraphics[scale=0.4]{./Jaxodraw/FermionPropFlowh} & m\includegraphics[scale=0.4]{./Jaxodraw/KroneckerDeltaUndotted}
  \end{pmatrix}\;.
\end{equation}
Like the fermion-fermion-scalar vertex (\eqref{eq:Sff}),
the appearance of a single dotted or undotted line
may imply a sign flip when reversing chirality-flow arrow directions 
(see below in \secref{sec:arrows and signs}).

Next, we consider the vector propagator for a massive particle which ---
aside from the trivial addition of a term $-m^2$ in the denominator ---
is unchanged
\begin{equation}
  \raisebox{-0.25\height}{\includegraphics[scale=0.45]{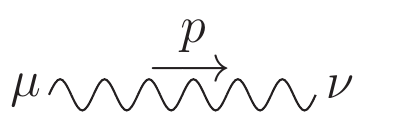}}
   \!\!\!\!\ =\;\; 
  \frac{-ig_{\mu\nu}}{p^2-m^2} \;\;
  \rightarrow\;\;\;
  -\frac{i}{p^2-m^2}\raisebox{-0.25\height}{\includegraphics[scale=0.45]{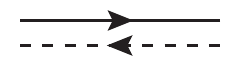}}
  \quad \mbox{or}\quad
  -\frac{i}{p^2-m^2}\raisebox{-0.25\height}{\includegraphics[scale=0.45]{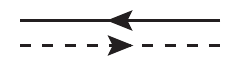}}\;.
  \label{eq:vbpropagator}
\end{equation}
Note that, as in the massless case, the arrow direction consistent with the
rest of the chirality-flow diagram must be chosen.
To complete the Standard Model list of propagators, we also need to treat
the scalar propagator, 
\begin{equation}
 \raisebox{-0.15\height}{\includegraphics[scale=0.45]{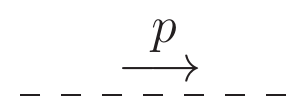}}
 \; =\;\; 
 \frac{i}{p^2-m^2}~,
\end{equation}
for which there is no flow of chirality, and hence no graphical representation
in the chirality-flow picture.

\subsection{Chirality-flow arrows and signs}
\label{sec:arrows and signs}

Now that we have the chirality-flow Feynman rules, we describe how
to apply them consistently.
For chirality-flow to work a continuous flow is required,
i.e., we cannot connect chirality-flow lines with arrows opposing
or pointing away from each other.
However, aligned arrows are not always immediately obtained.
When we connect two fermion lines in the chirality-flow formalism,
we often find situations where the chirality-flow arrows as stated in \secref{sec:diracspinors} 
point towards or away from each other, 
with (assuming massless, outgoing particles)
\begin{align}
\raisebox{-0.35\height}{\includegraphics[scale=0.45]{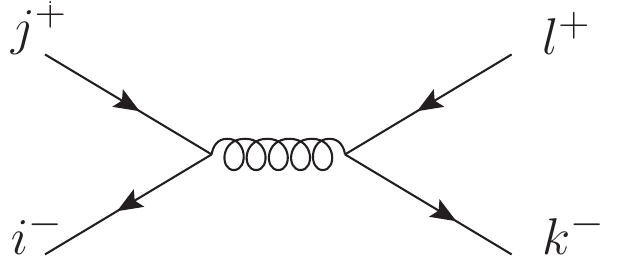}}
\sim
\underbrace{\raisebox{-0.35\height}{\includegraphics[scale=0.45]{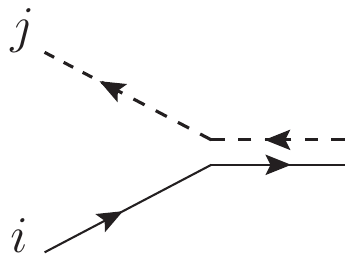}}}_{
\langle i|\taubar^\mu|j] }
\times
\underbrace{\raisebox{-0.35\height}{\includegraphics[scale=0.45]{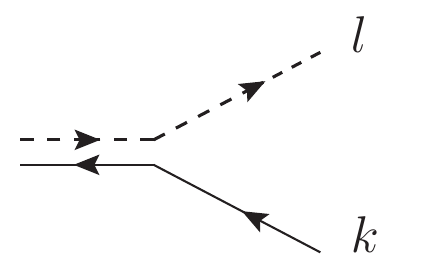}}}_{
\langle k|\taubar_\mu|l ]}~,
\label{eq:arrows opposed}
\end{align}
as the simplest example.
In such situations,
it is necessary to flip the chirality-flow arrows for one of the involved
fermion lines, above either to the left or to the right of the $\times$-sign.

In massless QED and QCD, each fermion line contains an odd number of
$\tau/\taubar$-matrices (one for each vertex and propagator).
Therefore, we can flip the chirality-flow arrow directions of a fermion line
with $n$ gauge bosons attached via the relation
\begin{align}
\langle i | \taubar^{\mu_1} \tau^{\mu_2} \dots \taubar^{\mu_{2n-1}} | j ] = 
[j | \tau^{\mu_{2n-1}} \dots \taubar^{\mu_2} \tau^{\mu_{1}} | i \rangle ~, 
\label{eq:sqAngle arrow swap}
\end{align}
where every even-numbered $\tau/\taubar$-matrix will be contracted with a momentum from the propagator
(via \eqref{eq:p_contr}),
and every odd-numbered $\tau/\taubar$-matrix corresponds to a fermion-fermion-vector vertex
(see \eqref{eq:fermion_photon_vertex}).
Assuming that the fermion arrow points to the left,
and that the attached bosons 
have outgoing momenta,
this gives the graphical representation
\begin{align}
\raisebox{-0.25\height}{\includegraphics[scale=0.4]{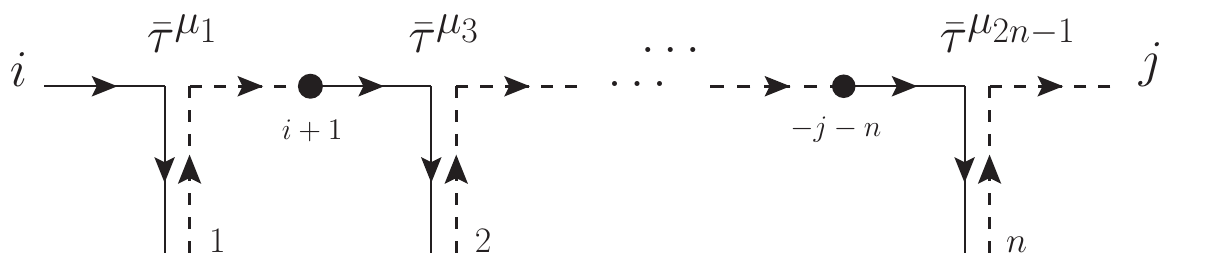}}
= 
\raisebox{-0.25\height}{\includegraphics[scale=0.4]{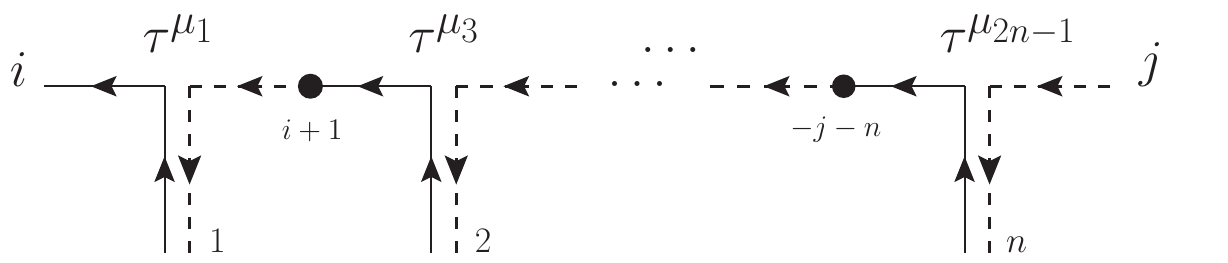}}~,
\label{eq:sqAngle arrow swap pic}
\end{align}
i.e., a swap of chirality-flow arrows (see sections 4.1 and 4.2 in \cite{Lifson:2020pai}).
Note that the labels of the momentum dots are unchanged by the arrow flip,
and that we have an {\it even} number of chirality-flow lines (consistent
with not obtaining an overall minus sign when changing all arrow directions).

We then connect different fermion lines using the chirality-flow structure of the vector propagator
\begin{align}
g_{\mu\nu} \leftrightarrow 
\raisebox{-0.25\height}{\includegraphics[scale=0.45]{./Jaxodraw/gmunua}}
\quad \text{or}
\quad \raisebox{-0.25\height}{\includegraphics[scale=0.45]{./Jaxodraw/gmunub}}~,
\label{eq:gmunu}
\end{align}
choosing the arrow directions which give a continuous flow.
For the simple example in \eqref{eq:arrows opposed} above,
we may for example flip the arrow directions on the left half and connect the flow lines to obtain
\begin{align}
\raisebox{-0.35\height}{\includegraphics[scale=0.45]{Jaxodraw/qqToQQFeyn}}
\sim
\underbrace{\raisebox{-0.35\height}{\includegraphics[scale=0.45]{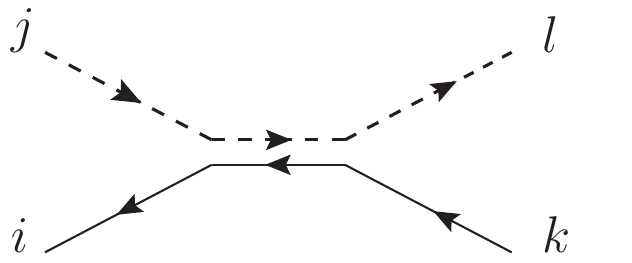}}}_{
\cSquare{jl}\cAngle{ki}}~.
\end{align}

The remaining two structures in massless QED and QCD, 
external momenta (\eqref{eq:Gordon p})
and external gauge bosons (\eqref{eq:massless pol vecs mu}),
both have the form of \eqref{eq:sqAngle arrow swap}
with $n=1$.
We conclude that Feynman diagrams in these theories only contain contractions of 
\eqref{eq:sqAngle arrow swap} \cite{Lifson:2020pai}.
Therefore,
for each disconnected piece\footnote{
  Disconnected pieces of chirality-flow diagrams are defined such that the arrow
  directions of two disconnected pieces do not affect each other. 
  They occur via non-Abelian vertices and internal scalars.
  For example, in \eqref{eq:VVV}
  the chirality-flow arrows of the double line ($g_{\mu_i\mu_j}$)
  are disconnected from the chirality-flow arrows of the momentum dot ($p_{\mu_k}$).
  \label{ft:disconnected}
} 
in a massless QED or QCD chirality-flow diagram, 
we only need to choose the chirality-flow arrow of one external particle in order to assign chirality-flow arrows.
Every other chirality-flow arrow connected to that particle by chirality-flow lines is then fixed by the vertex and propagator rules
(arrows flow through a vertex and fermion propagator,
and point in opposite directions in a boson double-line).

In contrast to massless QED and QCD, 
the full (tree-level) Standard Model also gives rise to fermions containing an even number of 
$\tau/\taubar$ matrices.
This is either due to the scalar-fermion-fermion vertex, \eqref{eq:Sff},
or due to the mass term of the fermion propagator, 
\eqref{eq:fermion prop pic},
each of which does not contain a $\tau/\taubar$.
To swap the chirality-flow arrow in such cases we use either of
\begin{align}
\langle i | \taubar^{\mu_1} \dots \tau^{\mu_{2n}} |  j\rangle &= 
- \langle j | \taubar^{\mu_{2n}} \dots \tau^{\mu_{1}} |  i\rangle~,
\nonumber \\
[i | \tau^{\mu_1} \dots \taubar^{\mu_{2n}} | j ] &= 
-[j | \tau^{\mu_{2n}} \dots \taubar^{\mu_{1}} |i ] ~, \label{eq:arrow swap even} 
\end{align}
each containing a minus sign.
This implies that signs now need to be tracked when swapping
chirality-flow arrows. We also remark that in the equation analogous to
\eqref{eq:sqAngle arrow swap pic}, we would now have an odd number of
chirality-flow lines.

\subsection{Application}
\label{sec:application}

With the above in mind,
we use the following method to calculate a Feynman diagram:
\begin{enumerate}
\item[(i)] Collect all common non-chirality-flow factors in front.
Such factors can come from vertices, propagators, 
and denominators of external gauge bosons.
\item[(iia)] If there are no fermion lines involved: 
\begin{itemize}
\item Without drawing the arrows,
draw and connect all chirality-flow lines according to the Feynman rules.
\item Since we have an even number of flow lines, we can choose 
an arbitrary chirality-flow arrow direction for one flow line.
Follow this arrow direction through the diagram, 
remembering that all double-lines should have opposing arrows,
and vertices and momentum dots should have a continuous flow.
\item Repeat for all disconnected chirality-flow pieces (see footnote \ref{ft:disconnected} for the definition of disconnected).
\end{itemize}
\item[(iib)] If there is at least one fermion line:
\begin{itemize}
\item For each fermion line and any bosons connected to it, 
draw the chirality-flow lines without arrows. 
Do not yet connect flow lines from different fermion lines.
\item Use \eqref{eq:lambdatildelower} 
to set the chirality-flow arrows of external fermions.
Then, 
for the rest of each fermion line draw the chirality-flow arrow directions in the only way possible.
\item Use \eqref{eq:sqAngle arrow swap} or 
(\ref{eq:arrow swap even})
to swap the chirality-flow arrows of each line, 
such that they can be connected with \eqref{eq:gmunu}.
Remember that we obtain a minus sign when swapping arrows on an odd number of flow lines.
\item For any remaining disconnected pieces, 
choose a chirality-flow direction for one particle and follow it
through the diagram as described in step (iia).
\end{itemize}
\end{enumerate}

\section{Examples}
\label{sec:examples}

In this section we illustrate some new features of the massive and
electroweak vertices, propagators and external particles. More extensive examples
of the massless formalism are given in section 6 in \cite{Lifson:2020pai}.
We remind the reader that thicker lines in Feynman diagrams imply that the particle is massive, 
that we use 't Hooft-Feynman gauge,
and that the Feynman rules are conveniently collected in \appref{sec:rosettastones_app}.

\subsection{\texorpdfstring{$e^+e^- \rightarrow  \gamma\gamma$}{e+e- -> gamma gamma}}
\label{sec:qqbTogg}

First we explore the effect of a massive fermion by
considering the Feynman diagram
\begin{equation}
    \includegraphics[scale=0.55,valign=c]{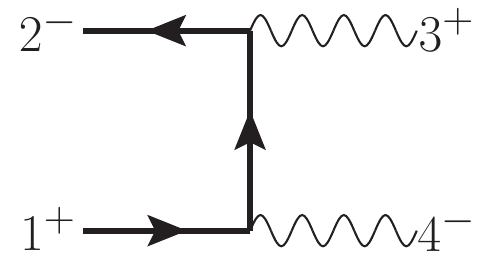}\;,
\end{equation}
corresponding to $e^-_{1+} e^+_{2-}\rightarrow \gamma_{3+} \gamma_{4-}$
where each particle is a helicity eigenstate.
In this example the effect of the mass enters at two levels, 
in the external particle wave functions and in the fermion propagator.

For the external spinors, 
we recall that (unlike in the massless case) 
a helicity eigenstate is a linear combination
of two different chirality eigenstates, as given in \secref{sec:helicity eigenstates}. 
Recalling 
\eqref{eq:uvbarpm_hel_cf}
we have 
\begin{equation}
   u^+(p_1)
  =
  \begin{pmatrix}
    -e^{\!-i\varphi_1}
    \raisebox{-5.5pt}{\includegraphics[scale=0.375]{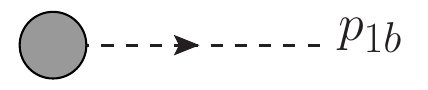}}
    \\
    \quad \quad \;\;\; \raisebox{-5.5pt}{\includegraphics[scale=0.375]{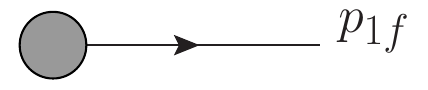}}
    \hspace{-1.75ex}
  \end{pmatrix}
  \quad \text{and} \quad  
  \bar{v}^-(p_2)
  =
  \bigg(
  \!
  \raisebox{-5.5pt}{\includegraphics[scale=0.375]{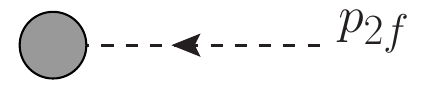}}
  \hspace{-2ex}
  \;\;,\;
  e^{i\varphi_2}
  \raisebox{-5.5pt}{\includegraphics[scale=0.375]{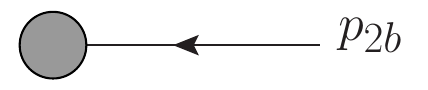}}\;
  \hspace{-2ex}
  \bigg)
  \;,
\end{equation}
where
$p^\mu_{i,\plus/\minus}=\frac{p_i^0 \pm|\vec{p}_i|}{2}(1,\pm\hat{p}_i)$ from \secref{sec:eigenvalue decomposition}. Explicit spinors can be found in \eqref{eq:pfb explicit},
and the phases are given by \eqref{eq:spinor phase def with alpha hel}
or (\ref{eq:spinor phase eigenvectors}).

Collecting overall factors from the vertices ($-ie\sqrt{2}$),
fermion propagator ($\frac{i}{(p_3-p_2)^2-m_e^2}$),
and two polarization vectors
($\frac{1}{\cAngle{r_3 3} \cSquare{4 r_4}}$, see 
\eqref{eq:polarisationBispinors+- massless}), 
then connecting the matching line types as given by
the Fermion propagator \eqref{eq:fermion prop pic},
we arrive at the expression
\begin{equation*}
\includegraphics[scale=0.55,valign=c]{Jaxodraw/exampleFeynFFVV} = (-ie\sqrt{2})^2 \frac{i}{(p_3-p_2)^2-m_e^2} \frac{1}{\cAngle{r_3 3} \cSquare{4 r_4}} \times
\end{equation*}
\begin{equation*}
  \left[\vphantom{ \includegraphics[scale=0.5,valign=c]{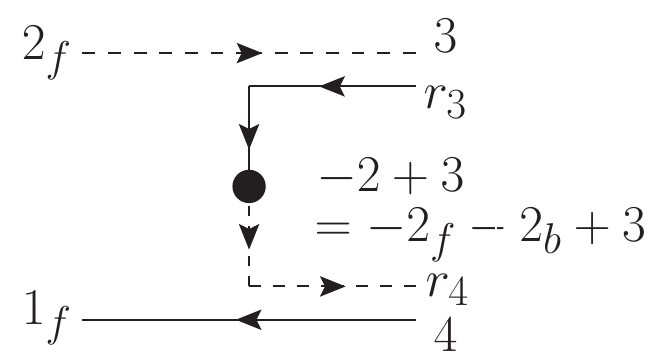}} \right.
  \underbrace{\includegraphics[scale=0.5,valign=c]{Jaxodraw/exampleMassiveChirFFVV1} }
    _{[2_f\,3] \langle r_3 | \Big(
      -|2_\plus \rangle [ 2_\plus| - |2_\minus \rangle [ 2_\minus|+ |3 \rangle [ 3|
          \Big) |r_4]\langle 4 \,1_\plus \rangle        }
        -e^{i(\varphi_2-\varphi_1)} \underbrace{\includegraphics[scale=0.5,valign=c]{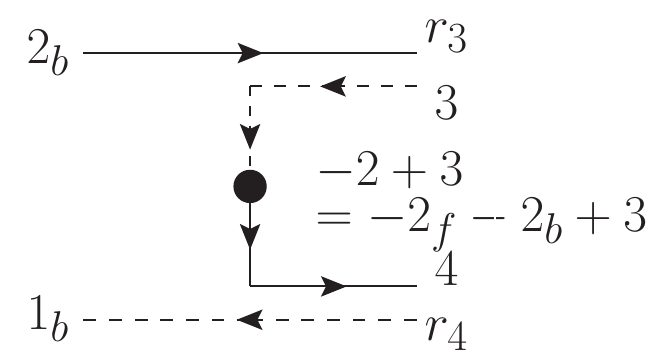}}_
        {\cAngle{2_\minus r_3} [3 \vert \Big( -\vert 2_\plus ] \langle 2_\plus \vert - \vert 2_\minus ] \langle 2_\minus \vert + \vert 3 ] \langle 3 \vert \Big) \vert 4 \rangle \cSquare{r_4 1_\minus}} 
\end{equation*}
\begin{equation}
   -m_e e^{-i\varphi_1}
  \underbrace{\includegraphics[scale=0.5,valign=c]{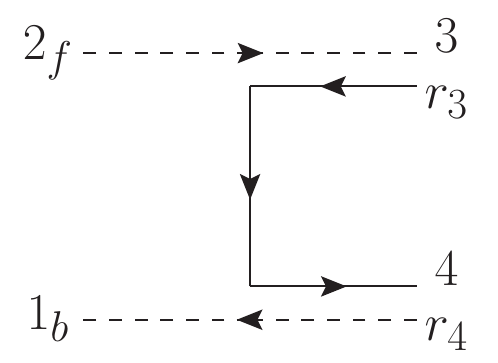}}_
             {\cSquare{2_\plus 3} \cAngle{r_3 4} \cSquare{r_4 1_\minus}} 
             + m_e e^{i\varphi_2} \underbrace{ \includegraphics[scale=0.5,valign=c]{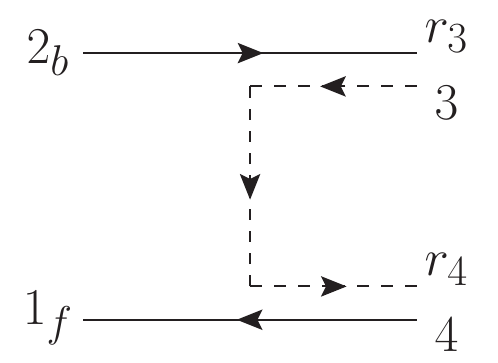}}_
       {\cAngle{2_\minus r_3} \cSquare{3 r_4} \cAngle{4 1_\plus}}
  \left. \vphantom{ \includegraphics[scale=0.5,valign=c]{Jaxodraw/exampleMassiveChirFFVV1}} \right]
       ~.
\end{equation}
While the line contractions themselves, via \eqref{eq:inner products}, give the
diagram expressed in terms of spinor inner products, we have written out the
contractions in terms of brackets as a service to the reader.

We note that when the mass goes to zero, both Weyl spinors $|1_\minus]$ and $\langle 2_\minus |$ also go to zero as
they are proportional to $\sqrt{\lambda_{i,\minus}} = \sqrt{p_i^0 - \vert \vec{p}_i \vert}$.
This implies that the only surviving chirality-flow diagram in the massless case
is the first diagram above, (cf. eq.\,(6.19) in \cite{Lifson:2020pai}).
We further note that in the massless limit this Feynman diagram can be removed by an unphysical
gauge choice, using $r_4=p_3$ and $r_3=p_2= p_{2f}$ 
(implying that the diagram with the photons exchanged cannot be removed by this gauge choice).

We also remark that for this diagram, we did not need to adjust any chirality-flow
arrows since we only have one fermion line, and the arrows from the
external photon polarization vectors can be aligned accordingly, without additional
minus signs.

\subsection{\texorpdfstring{$e^+e^- \rightarrow Zh$}{e+e- -> Zh} }
\label{sec:AssocHiggs}

Next we consider a diagram relevant for associated Higgs production,
$e^-_{1+}e^+_{2-}\rightarrow h_3 Z_{4,0}$, 

\begin{equation}
  \includegraphics[scale=0.55,valign=c]{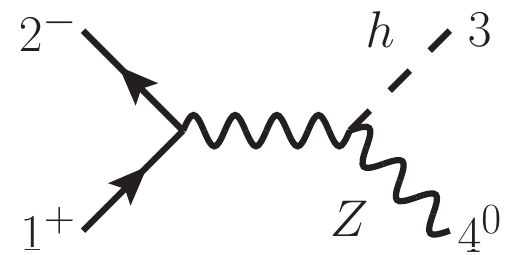}\;.
    \label{eq:testname}
\end{equation}
Aside from a massive fermion, this diagram involves both an internal and an
external massive vector boson, the vector-vector-scalar vertex from \eqref{eq:VVS},
and a $Z$-fermion-fermion vertex with different couplings to left- and right-chiral
particles.
For the $Z$-boson, the superscript 0 in $4^0$ refers to the longitudinal
polarization, with chirality-flow representation given in \eqref{eq:eps0 dot}.
We will use the general spin basis,
such that the spins are measured relative to spin axes $\smu_i$,
\textit{not} relative to the particles' directions of motion.

Collecting overall factors from the $Zff$- and $ZZh$-vertices
(\eqsrefa{eq:fermion_photon_vertex}{eq:VVS}), 
from the massive spin-1 propagator (\eqref{eq:vbpropagator}), 
as well as from the normalization of the polarization vector (\eqref{eq:eps0 dot}), 
and treating the massive external fermions using the general spin decomposition 
\eqref{eq:uvbarpm_spin_cf} 
\begin{eqnarray}
  u^+(p_1)
  &=&
  \begin{pmatrix}
    - e^{\!-i\varphi_1}
    \!\sqrt{\alpha_1}
    \raisebox{-5.5pt}{\includegraphics[scale=0.375]{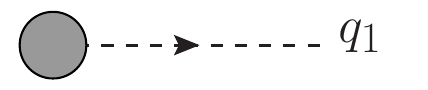}}
    \hspace{-1.75ex}
    \\
    \phantom{- e^{\!-i\varphi_1}
    \!\sqrt{\alpha_1}}
    \raisebox{-5.5pt}{\includegraphics[scale=0.375]{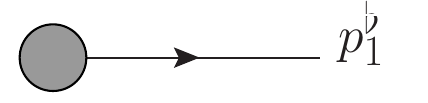}}
    \hspace{-1.75ex}
  \end{pmatrix}~,
\nonumber\\
    \bar{v}^-(p_2)
  &=&
  \bigg(
  \!
  \raisebox{-5.5pt}{\includegraphics[scale=0.375]{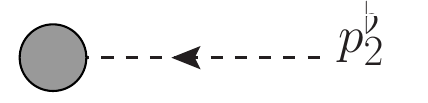}}
  \hspace{-2ex}
  \;,\;
  e^{i\varphi_2}\!
  \sqrt{\alpha_2}
  \raisebox{-5.5pt}{\includegraphics[scale=0.375]{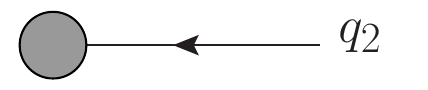}}
  \hspace{-2ex}
  \bigg)~,
\end{eqnarray}
where the lightlike vectors $p^\flat$ and $q$ are given by \eqref{eq:lightlikevectordecomposition},
we get two chirality-flow diagrams
\begin{equation*}
  \hspace*{-3.3cm}\includegraphics[scale=0.55,valign=c]{Jaxodraw/exampleMassiveFeynFFZH}
  = ie\sqrt{2}\frac{-i}{(p_1+p_2)^2-m_Z^2}ie\frac{m_Z}{\sin \theta_W \cos\theta_W}\frac{1}{m_Z\sqrt{2}} \times
\end{equation*}
\begin{align}
  \left[\vphantom{ \includegraphics[scale=0.5,valign=c]{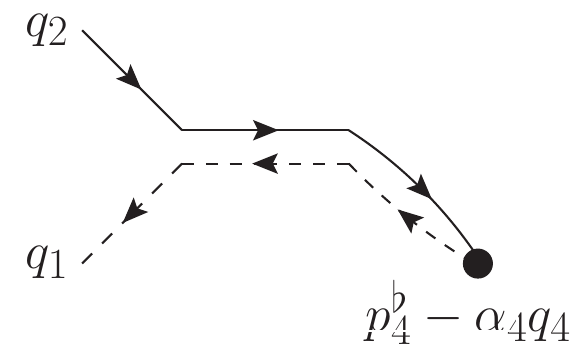}} \right.
    -C_L\sqrt{\al_1\al_2}  e^{i(\varphi_2-\varphi_1)}  \underbrace{\includegraphics[scale=0.5,valign=c]{Jaxodraw/exampleMassiveChirLeftFFZH}}_{
     \langle q_2 \vert \Big( \vert p_4^\flat \rangle [ p_4^\flat \vert - \alpha_4 \vert q_4 \rangle [ q_4 \vert \Big) \vert q_1 ]}
      + \
       C_R\underbrace{ \includegraphics[scale=0.5,valign=c]{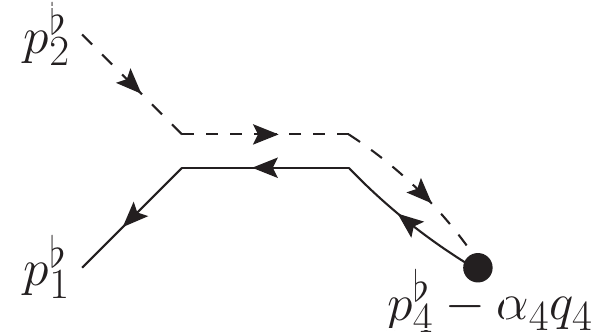}}_{
      [p^\flat_2 \vert \Big( \vert p_4^\flat ] \langle p_4^\flat \vert - \alpha_ 4 \vert q_4 ] \langle q_4 \vert \Big) \vert p^\flat_1 \rangle}
  \left.\vphantom{\includegraphics[scale=0.5,valign=c]{Jaxodraw/exampleMassiveChirLeftFFZH}}\right]\;.
\end{align}
Explicit Weyl spinors can be found in \eqref{eq:massless Weyl explicit}
and the phases $\varphi_{1,2}$ are given
by \eqref{eq:spinor phase def with alpha}.

In the Standard Model, the couplings $C_L,C_R$ 
assume the values given below \eqref{eq:fermion_photon_vertex},
however, here they can be adjusted to match the theory at hand.
As in the previous example, we draw the chirality-flow arrows as dictated
by the external fermions, but note that flipping all of the arrows does
not introduce any additional signs.

We see that only the second chirality-flow diagram contributes when the fermions are massless since 
$\alpha_{1,2}\rightarrow 0$.
Further, we can simplify this calculation with a clever choice of $q_{1,2,4}$. 
For instance, we can remove all but one of the four terms with the choice
$q_1 = p_4^\flat$, 
$q_2 = q_4 = p_1^\flat$, 
which fixes the axes $\smu_i$ 
along which the spin of each particle is measured for all Feynman diagrams contributing to this process.

\subsection{\texorpdfstring{$ q\bar{q} \rightarrow q\bar{q}h$}{qq -> qqh}}
\label{sec:4fh}

As a final example we consider $q_{1-}\bar{q}_{2-} \rightarrow q_{3+} \bar{q}_{4+} h_5$,
and calculate
\begin{equation}
    \includegraphics[scale=0.5,valign=c]{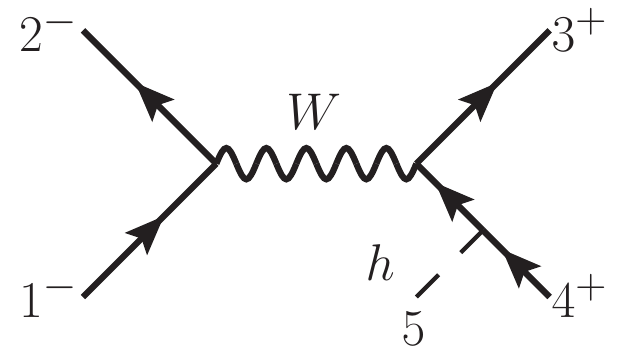}~,
\end{equation}
which nicely displays the chiral nature of the $W$-boson, and demonstrates
the treatment of chirality-flow arrow signs when multiple fermion lines are present.
Note that if we use massless spinors (but keep the mass in the coupling),
then the corresponding helicity amplitude will vanish since the incoming particles
have the same helicity.

We again begin by collecting overall factors from the vertices
(\eqsrefa{eq:fermion_photon_vertex}{eq:Sff}) 
and the two propagators (\eqsrefa{eq:vbpropagator}{eq:fermion prop pic}).
We also include Kronecker deltas in color for the quarks, $\delta_{{q_2}{q_1}}\delta_{{q_3}{q_4}}$.
The massive fermions are treated using the general spin decomposition 
(\eqsrefa{eq:ubarvpm_spin_cf}{eq:uvbarpm_spin_cf}),
but we do \textit{not} immediately join them together with the vector propagator,
\begin{eqnarray}
  &&\includegraphics[scale=0.5,valign=c]{Jaxodraw/exampleMassiveFeynFFFFH} = (ie\sqrt{2})^2 \frac{-i}{(p_1+p_2)^2-m_W^2}
   \frac{i}{(p_4+p_5)^2-m_4^2}ie\left( -\frac{m_4}{2 \sin \theta_W m_W} \right) \delta_{{q_2}{q_1}}\delta_{{q_3}{q_4}} \nonumber\\
    && \times  
   C_{L,12} \sqrt{\al_2}e^{i\varphi_2}\hspace*{-0.05 cm}
   \includegraphics[scale=0.45,valign=c]{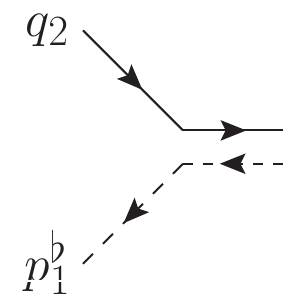} 
    \times C_{L,34}\sqrt{\al_3} (-e^{i\varphi_3})\left[ \sqrt{\al_4} (-e^{i\varphi_4}) \includegraphics[scale=0.45,valign=c]{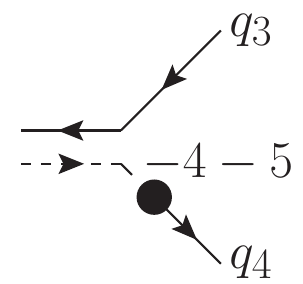} 
    +  m_4 \includegraphics[scale=0.45,valign=c]{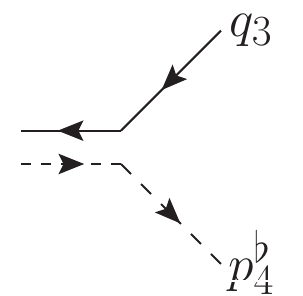} \right].
    \label{eq:W example factorised}
\end{eqnarray}
Here, the internal, unlabeled ends of chirality-flow lines 
are not yet connected to external spinors,
and the couplings are assumed to take their Standard Model values given below 
\eqref{eq:fermion_photon_vertex}, 
but can be changed to suit the theory at hand.

On the solid (dotted) lines, to be joined in the middle,
the chirality-flow arrows are pointing towards (away from) each other.
To join the flow lines, 
we must either flip the flow arrows on the first term 
(containing $q_2,p_1^\flat$)
or on the second (in square brackets). 
If we flip them on the first term we use 
\begin{equation}
    \includegraphics[scale=0.5,valign=c]{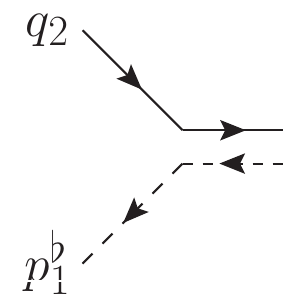} = 
    \includegraphics[scale=0.5,valign=c]{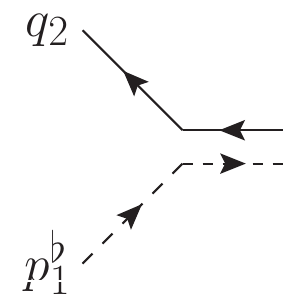}, \label{eq:sqAngle arrow swap example}
\end{equation}
a special case of \eqref{eq:sqAngle arrow swap pic}.
Note that this arrow swap is not possible at the vertex level itself 
(i.e.\ $\tau \neq \taubar$),
but requires each external flow line to be contracted with a Weyl spinor.

The Feynman diagram is then immediately written down in terms of inner products
\begin{eqnarray}
  &&\includegraphics[scale=0.5,valign=c]{Jaxodraw/exampleMassiveFeynFFFFH}  =\nonumber \\
  &&  (ie\sqrt{2})^2 \frac{-i}{(p_1+p_2)^2-m_W^2} \frac{i}{(p_4+p_5)^2-m_4^2}ie
    \left( -\frac{m_4}{2 \sin \theta_W m_W} \right)\delta_{{q_2}{q_1}} \delta_{{q_3}{q_4}} 
     C_{L,12} C_{L,34} \sqrt{\al_2\al_3} e^{i(\varphi_2+\varphi_3)} \nonumber \\
    &&\times \left[\vphantom{ \includegraphics[scale=0.5,valign=c]{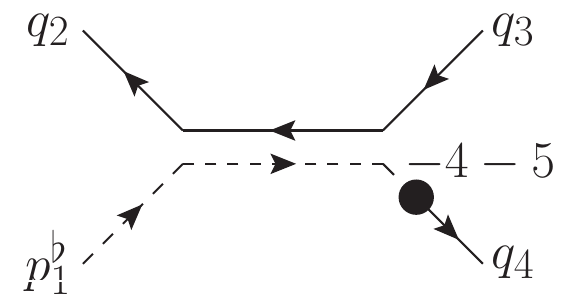}} \right.
    \sqrt{\al_4} e^{i\varphi_4}\underbrace{\includegraphics[scale=0.45,valign=c]{Jaxodraw/exampleMassiveChirFFFFHfullDot} }_{
    \cAngle{q_3 q_2} [p_1^\flat | \Big( - | p_4^\flat ] \langle p_4^\flat| - \al_4 | q_4 ] \langle q_4|-| p_5^\flat ] \langle p_5^\flat| - \al_5 | q_5 ] \langle q_5|  \Big) |q_4 \rangle} -  
    m_4 \underbrace{ \includegraphics[scale=0.45,valign=c]{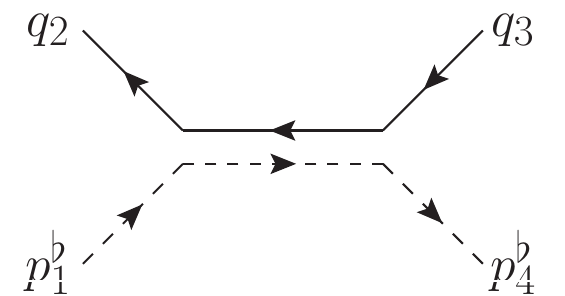}}_{
     \cAngle{q_3 q_2} \cSquare{p_1^\flat p_4^\flat}}~
    \left. \vphantom{ \includegraphics[scale=0.5,valign=c]{Jaxodraw/exampleMassiveChirFFFFHfullDot}} \right],
   \label{eq:FFFH answer}
\end{eqnarray}
which vanishes in the massless limit 
($\al_i \rightarrow 0$) as it should.
We can also make this diagram disappear by choosing $q_2=q_3$,
allowing to simplify the amplitude calculation.

Note that we could alternatively have flipped the arrows on the chirality-flow factors in the square bracket of 
\eqref{eq:W example factorised}.
In this case we must be more careful,
since the term proportional to $\sqrt{\al_4}$ comes with a minus sign under arrow reversal,
\begin{equation}
    \includegraphics[scale=0.5,valign=c]{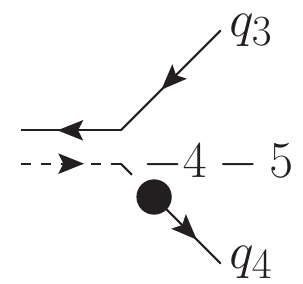} = 
    -  \includegraphics[scale=0.5,valign=c]{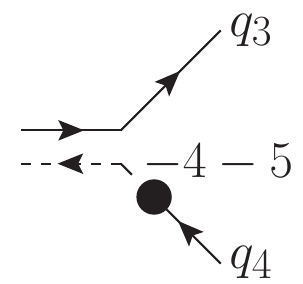}, \label{eq:angleAngle arrow swap example}
\end{equation}
a special case of \eqref{eq:arrow swap even}.
The chirality-flow graphs in \eqref{eq:FFFH answer} would then be written down with the arrow direction reversed,
giving a factor $(-1)^3$
in the first graph to compensate the minus sign in
\eqref{eq:angleAngle arrow swap example}.
Thus, 
in accordance with \secref{sec:arrows and signs},
we obtain the same result regardless of which fermion line had its chirality-flow arrows flipped.

\section{Conclusion and outlook}
\label{sec:conclusion}

In a recent paper we showed that it is possible to define a chirality-flow
description of the Lorentz structure of Feynman rules and diagrams for massless QED and QCD.
Using this method, along with
its graphical interpretation, the values of amplitudes from Feynman diagrams can
be immediately written down in a transparent and intuitive manner,
offering superior simplicity for helicity-assigned diagrams.

Here we prove that the flow idea can be extended to the full standard
model, 
in essence by treating massive fermions and spin-1 bosons as
combinations of their massless analogs.
In principle, this extension is to be anticipated, since the ``flows'' describing the
Lorentz structure are nothing but the only Lorentz-invariant quantities
at hand, the antisymmetric contraction of two left- or right-chiral spinors, i.e.,
the spinor inner products.
For the same reason, we expect the formulation of Feynman rules
beyond leading order, or for physics beyond the Standard Model to be straightforward.

The addition of mass implies a modest complication, resulting in twice as
many terms for external fermions, as well as for internal Fermion propagators,
and another polarization vector to keep track of for external vector bosons.
On the other hand, the chirality-flow diagrams are simplified for weak interactions
involving $W$-bosons, where --- due to the left chiral nature --- only one
term in the vector-fermion vertex is present.

Aside from offering an intuitive and transparent way of thinking about and
calculating Feynman diagrams, we expect our method to be computationally
beneficial (at least) for programs that base their calculations on
helicity-assigned Feynman diagrams,
due to the complete avoidance of Dirac matrices.

\acknowledgments 
We thank Johan Bijnens, Rikkert Frederix  
and Simon Pl\"atzer
for constructive feedback on the manuscript.
This work was supported by 
the Swedish Research Council (contract no.\ 2016-05996),
as well as the European Union's Horizon 2020 research and innovation programme
(grant agreement no.\ 668679).
This work has also received funding from the European Union's Horizon 2020 research and innovation programme as part of the Marie Sklodowska-Curie Innovative Training Network MCnetITN3 (grant agreement no.\ 722104).

\appendix

\section{Dirac spinors}
\label{sec:diracspinors_app}

\subsection{Conventions and the chiral representation}
\label{sec:diracspinors_properties_app}
Here we summarize our conventions for spinors and gamma matrices in the chiral basis. 
For a more complete set of our conventions, 
as well as some useful algebraic relations see \cite{Lifson:2020pai}.
For general reviews on chiral spinors and the spinor-helicity formalism we refer to \cite{Mangano:1990by,Dixon:1996wi,Dittmaier:1998nn,Weinzierl:2005dd,Weinzierl:2007vk,Dreiner:2008tw,Ellis:2011cr,Peskin:2011in,Elvang:2013cua,Dixon:2013uaa}.

We use the Dirac matrices in the chiral, 
or Weyl basis,
\begin{align}
\gamma^{\mu} 
= 
\begin{pmatrix}
0 & \sigma^{\mu} \\
\Bar{\sigma}^{\mu} & 0
\end{pmatrix}
=
\begin{pmatrix}
0 & \sqrt{2}\tau^{\mu} \\
\sqrt{2}\Bar{\tau}^{\mu} & 0
\end{pmatrix}
\;,
\quad
\gamma^5 
= 
i \gamma^0 \gamma^1 \gamma^2\gamma^3 
=
\begin{pmatrix}
-1 & 0 \\
0 & 1
\end{pmatrix}
\;,
\end{align}
where the $\sigma$-matrices are the Pauli matrices,
\begin{align}
\sigma^\mu
=
(\sigma^0,\vec{\sigma})
=
(\sigma^0,\sigma^1,\sigma^2,\sigma^3)
=
\left(
\begin{pmatrix}
1 & 0\\
0 & 1 
\end{pmatrix}
\,,\,
\begin{pmatrix}
0 & 1\\
1 & 0 
\end{pmatrix}
\,,\,
\begin{pmatrix}
0 & -i\\
i & 0 
\end{pmatrix}
\,,\,
\begin{pmatrix}
1 & 0\\
0 & -1 
\end{pmatrix}\right)
~, \quad
\bar{\sigma}^\mu
=
(\sigma^0,-\vec{\sigma})
\;,
\end{align}
and the $\tau$-matrices (Infeld-van-der-Waerden matrices) are
$\tau^\mu=\frac{1}{\sqrt{2}}\sigma^\mu$ and $\taubar^\mu=\frac{1}{\sqrt{2}}\sigmabar^\mu$. 
The normalization of the $\tau$-matrices is chosen such that no unnecessary powers of $2$ are carried around in the algebraic relations,
\begin{align}
\label{eq:tautrace}
\trace\big(\tau^{\mu}\taubar^\nu\big)
&=g^{\mu\nu}\, 
\quad \Leftrightarrow \quad
\trace\big(\taubar^\mu\taubar^\nu\big)
= \trace\big(\tau^\mu\tau^\nu\big)
= \delta^{\mu\nu}\,,\\
\label{eq:taufierz}
\tau_\mu^{\dal\be}\taubar^\mu_{\ga\deta}
&=\delta_{\ga}^{\ \be}\delta_{\ \deta}^{\dal}\,, \quad
\taubar^\mu_{\al\dbe}\taubar_{\mu,\ga\deta}
=\eps_{\al\ga}\eps_{\dbe\deta}\,,\quad
\tau^{\mu,\dal\be}\tau_{\mu}^{\dga\eta}
=\eps^{\dal\dga}\eps^{\be\eta}\,,
\end{align}
where 
$g^{\mu\nu}=\mathrm{diag}(1,-1,-1,-1)$ 
denotes the Minkowski metric and $\eps$ the Levi-Civita tensor defined in 
\eqref{eq:epsilon_action}.
Note that $\taubar^\mu$ has lower indices, with the first index undotted,
while $\tau^\mu$ has upper indices with the first index dotted.

Since we use the chiral basis,
the momentum-space Dirac spinors divide into their left- 
and right-chiral parts which can be projected out using the projection operator
$P_{R/L} = \frac{1}{2}(1\pm\ga^5)$ 
\begin{align}
u^{\Js}
= 
\begin{pmatrix} 
|u^{\Js}_L] 
\\ 
|u^{\Js}_R\rangle
\end{pmatrix}
\quad\text{and}\quad
v^{\Js}
= 
\begin{pmatrix} 
|v^{\Js}_L] 
\\ 
|v^{\Js}_R\rangle 
\end{pmatrix}
\;.\quad
\end{align}
Here, $\Js$ is the spin along axis $\smu$ 
(see \appref{sec:massive_spin_operator_app}),
and the $L/R$ 
labels refer to particles which transform under the left- 
and right-chiral representations of the (restricted) Lorentz group.
(Compared to \secref{sec:diracspinors} we have absorbed a factor
$\sim \sqrt{\alpha} e^{\pm i \varphi}$ into the kets as required.)

\subsection{Relativistic spin operator for massive spinors}
\label{sec:massive_spin_operator_app}

The threevector spin operator $\Si^i/2$ 
can be promoted to a fourvector operator $\Si^\mu/2$ (see e.g.\ \cite{Ohlsson:2011zz})
defined as\footnote{
Note that the spin operator $\Si^\mu$ 
is directly related to the Pauli-Lubanski operator $W^\mu$ 
via $\Si^\mu = \frac{2}{m}W^\mu$.
We remind that 
$W^2 = -m^2J(J+1)$
is one of the two quadratic Casimirs of the Poincar\'e algebra
(the other being $P^2=m^2$), 
where $J$ is the total spin.
}
\begin{equation} \label{eq:Sigma 4vec defn}
\frac{1}{2}\Si^\mu = -\frac{1}{4m}\eps^{\mu\nu\la\omega}P_\nu\si_{\la\omega}~,
\end{equation}
where $P_{\nu}$ is the momentum operator ($P_{\nu} = i\partial/\partial x^{\nu}$),
$\eps^{\mu\nu\la\omega}$ ($\eps^{0123}\defequal 1 \Rightarrow \eps_{0123}= -1$) 
is the four-dimensional Levi-Civita tensor,
and $\si^{\mu\nu}$ is defined as
\begin{equation} \label{eq:sigma mu nu}
\si^{\mu\nu} = \frac{i}{2}\left[\ga^\mu,\ga^\nu\right]~.
\end{equation}
In the rest frame of the particle ($p^\mu \eqAtRest (m,\vec{0})$), 
\eqref{eq:Sigma 4vec defn} 
takes the form of the familiar spin operator
\begin{align} \label{eq:Sigma defn}
\frac{1}{2}\Si^i \eqAtRest 
 \frac{i}{4}\eps^{ijk}\ga^j\ga^k = \frac{1}{2}\ga^5\ga^0\ga^i
=
\frac{1}{2}
\begin{pmatrix}
\si^i & 0 \\
0 & \si^i
\end{pmatrix}~,
\qquad 
\frac{1}{2}\Si^0 \eqAtRest 0~,
\end{align}
which also shows that $\Si^\mu p_\mu = 0$.

For massive particles, 
we measure the total spin $J$ 
and its spin $\Js$ projected onto some axis $\smu$ 
which must equal 
(see e.g.\ \cite{Dreiner:2008tw,Ohlsson:2011zz})
\begin{equation}
\smu \eqAtRest (0,\sAxisHat)~, 
\end{equation}
in the rest frame. 
From this frame, we also find the relations
\begin{equation}
\qquad \sAxisSq = -1~, \qquad \sDotp = 0~,
\end{equation}
and that for Dirac spinors,
the spin projected onto $\smu$ is given by the operator
\begin{equation} \label{eq:spin on motion rest}
 \mathcal{O}_s=-\frac{\Si^\mu \smuLow}{2} =
 \frac{1}{4m}\eps^{\mu\nu\la\omega}\smuLow P_\nu\si_{\la\omega}~.
\end{equation}

To obtain the last form of  $\mathcal{O}_s$ in \eqref{eq:Js spinors} 
we use the identity
\begin{align} \label{eq:ga5 sigma mu nu identity}
\frac{1}{2}\eps^{\mu\nu\la\omega}\si_{\la\omega} = -i\ga^5\si^{\mu\nu} = \ga^5 (\ga^\mu\ga^\nu - g^{\mu\nu})~,
\end{align}
to find
\begin{align} \label{eq:spin projected operator}
\mathcal{O}_s = -\frac{\Si^\mu s_\mu}{2} = \frac{1}{2m}\ga^5\left(\ga^\mu\ga^\nu - g^{\mu\nu}\right)P_\nu s_\mu 
= \frac{1}{2m}\ga^5\slashed{s}\slashed{P}~,
\end{align}
where we used that $\sDotp = 0$.
We act $P_\mu \equiv i\partial_\mu$ on the $u$ and $v$ spinors,
then make use of the Dirac equation $(i\slashed{\partial}-m)\psi=0$
to
replace the operator $\slashed{P}$ 
with $m$ for any on-shell spinor.

Making this substitution in \eqref{eq:spin projected operator}
gives \eqref{eq:Js spinors},
namely
\begin{align}
  \mathcal{O}_s=
  -\frac{\Si^\mu s_\mu}{2} 
  = \frac{1}{2}\ga^5\smu\ga_\mu~.
\end{align}

\section{Weyl spinors}
\label{sec:weyl spinors app}
Here we give some useful properties and explicit values of the Weyl spinors and their inner products.
 
\subsection{Explicit representations of spinors and their inner products}
\label{sec:spinor reps app}

A generic massless Weyl spinor of (real) momentum $p$ 
can be expressed in terms of light-cone coordinates
$p^{\pm} = p^0\pm p^3$, $p^{\perp} = p^1+ip^2$\ 
and $p^{\perp^*}=p^1-ip^2$
(see appendix A.2 of \cite{Lifson:2020pai} for more details)
\begin{alignat}{2}
|p\rangle
&=
\frac{e^{-i\theta/2}}{\sqrt{|p^+|}}
\begin{pmatrix}p^+\\p^{\perp}\end{pmatrix}
\;,
\qquad
&&|p]
=\mathrm{sgn}(p^0)
\frac{e^{i\theta/2}}{\sqrt{|p^+|}}
\begin{pmatrix}p^{\perp^*}\\-p^+\end{pmatrix}~,
\nonumber \\
\langle p|
&=
\frac{e^{-i\theta/2}}{\sqrt{|p^+|}}
\Big(\,p^{\perp}\,,\,-p^+\,\Big)
\;,
\qquad
&&[p|
=\mathrm{sgn}(p^0)
\frac{e^{i\theta/2}}{\sqrt{|p^+|}}
\Big(\,p^+\,,\,p^{\perp^*}\,\Big)
\;, 
\label{eq:massless Weyl explicit}
\end{alignat}
where the phase $\theta$ is a little group phase which we set to $0$ 
throughout the rest of the paper,
and the $\mathrm{sgn}(p^0)$ 
term is required\footnote{
This factor $\mathrm{sgn}(p^0)$ 
can alternatively be placed on the angled spinors
or even be distributed between the square and angle spinors,
as long as \eqref{eq:mom bispinor -p} holds. }
to consistently take $p \rightarrow -p$ in
\eqref{eq:Mom two-spinors}
\begin{align}
\slashed{p} &\eqForPosE |p]\langle p| 
\rightarrow
(-\slashed{p})  
= |-p]\langle -p|  
= - |p]\langle p| ~,
\qquad 
 \bar{\slashed{p}} \eqForPosE |p\rangle [p|
\rightarrow
  (-\bar{\slashed{p}}) 
= |-p\rangle [-p|
  = -|p\rangle [p| 
~. \label{eq:mom bispinor -p}
\end{align}
In the special frame $p^+ = p^\perp = 0$
a valid representation of the Weyl spinor is
\begin{alignat}{2}
|p\rangle
&=
\begin{pmatrix}0 \\ \sqrt{|p^-|}\end{pmatrix}
\;,
\qquad
&&|p]
=\mathrm{sgn}(p^0)
\begin{pmatrix} \sqrt{|p^-|} \\ 0 \end{pmatrix}~,
\nonumber \\
\langle p|
&=
\Big(\,\sqrt{|p^-|}\,,\,0\,\Big)
\;,
\qquad
&&[p|
=\mathrm{sgn}(p^0)
\Big(\,0\,,\,\sqrt{|p^-|}\,\Big)
\;,
\label{eq:massless Weyl explicit p^+ = 0}
\end{alignat}
which affects \eqsrefa{eq:explicitproducts}{eq:phase in pf and pb} below.

From \eqref{eq:massless Weyl explicit} (with $\theta=0$)
we obtain the explicit forms of the inner products
\begin{equation}
\langle i j \rangle 
= 
\frac{1}{\sqrt{|p_i^+||p_j^+|}}\big(p_i^\perp p_j^+ - p_j^\perp p_i^+\big)
~~,\quad 
[i j] 
= 
\frac{\mathrm{sgn}(p_i^0p_j^0)}{\sqrt{|p_i^+||p_j^+|}}\big(p_i^+ p_j^{\perp^*} - p_j^+p_i^{\perp^*}\big)~~,
\label{eq:explicitproducts}
\end{equation}
as well as the Hermitian conjugation relations
\begin{equation}
|p\rangle^\dagger = \mathrm{sgn}(p^0)[p|~, \qquad 
|p]^\dagger = \mathrm{sgn}(p^0)\langle p|~. \label{eq:Hermitian conjugation}
\end{equation}

Since \eqref{eq:massless Weyl explicit} holds for any massless spinor,
we can use it to represent both the spinors of momenta 
$p^\flat$ and $q$ introduced in \secref{sec:fourvectors},
or the forward and backward moving spinors of momenta 
$p_{\plus/\minus}$ introduced in \secref{sec:eigenvalue decomposition}
(for which the spinor wave functions are helicity eigenstates).
For convenience, 
we will also give the $p_{\plus/\minus}$ 
spinors with $p^0>0$ from 
\eqref{eq:spinor phase eigenvectors} in terms of $\phat = \vp/|\vp|$
\begin{alignat}{2}
\eigRanp{p}
&=
\sqrt{\frac{\eigvalp}{2(1+\phat^3)}}
\begin{pmatrix}1+\phat^3 \\ \phat^{\perp}\end{pmatrix}
\;,
\qquad
&&\eigSqRp{p}
=
\sqrt{\frac{\eigvalp}{2(1+\phat^3)}}
\begin{pmatrix}\phat^{\perp^*}\\-(1+\phat^3)\end{pmatrix}~,
\nonumber \\
\eigRanm{p}
&=
\sqrt{\frac{\eigvalm}{2(1-\phat^3)}}
\begin{pmatrix}1-\phat^3 \\ -\phat^{\perp}\end{pmatrix}
\;,
\qquad
&&\eigSqRm{p}
=
\sqrt{\frac{\eigvalm}{2(1-\phat^3)}}
\begin{pmatrix}-\phat^{\perp^*}\\-(1-\phat^3)\end{pmatrix}~,
\nonumber \\
\eigLanp{p}
&=
\sqrt{\frac{\eigvalp}{2(1+\phat^3)}}
\Big(\,\phat^{\perp}\,,\,-(1+\phat^3)\,\Big)
\;,
\qquad
&&\eigSqLp{p}
=
\sqrt{\frac{\eigvalp}{2(1+\phat^3)}}
\Big(\,1+\phat^3\,,\,\phat^{\perp^*}\,\Big)
\;,
\nonumber \\
\eigLanm{p}
&=
\sqrt{\frac{\eigvalm}{2(1-\phat^3)}}
\Big(-\phat^{\perp}\,,\,-(1-\phat^3)\,\Big)
\;,
\qquad
&&\eigSqLm{p}
=
\sqrt{\frac{\eigvalm}{2(1-\phat^3)}}
\Big(\,1-\phat^3\,,\,-\phat^{\perp^*}\,\Big)
\;.
\label{eq:pfb explicit}
\end{alignat}

\subsection{Useful Identities}
\label{sec:useful identities}
In addition to momentum conservation and judicious multiplication by 1 
it is often useful to use the Schouten identity\footnote{
  The Schouten identity follows from the fact that any three two-component spinors are linearly dependent.},
which has a nice flow representation
\begin{align} 
\underbrace{\includegraphics[scale=0.45,valign=c]{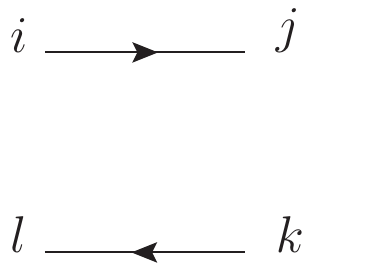}}_{
\langle ij\rangle\langle kl\rangle }
&=
\underbrace{\includegraphics[scale=0.45,valign=c]{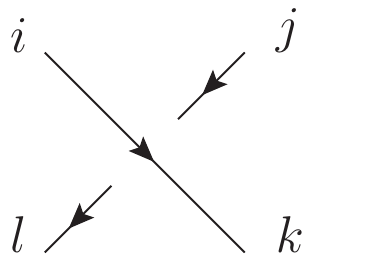}}_{
\langle ik\rangle\langle jl\rangle} 
+ 
\underbrace{\includegraphics[scale=0.45,valign=c]{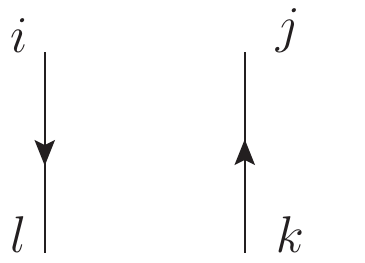}}_{
\langle il\rangle\langle kj\rangle }~,
\nonumber \\ 
\label{eq:Schouten2}
\,
\underbrace{\includegraphics[scale=0.45,valign=c]{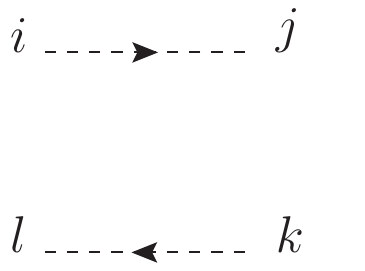}}_{
[ij]\,[kl]}
&=
\underbrace{\includegraphics[scale=0.45,valign=c]{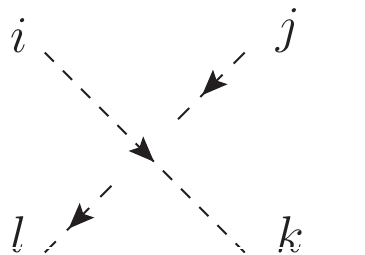}}_{
[ik]\,[jl]} 
+
\underbrace{\includegraphics[scale=0.45,valign=c]{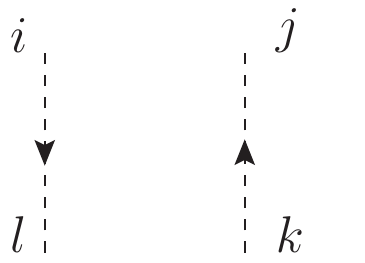}}_{
[il]\,[kj]} 
\,
~,
\end{align}
and the relation for $s_{ij}$
\begin{equation}
s_{ij} = 2p_i\cdot p_j = \cAngle{ij}\cSquare{ji}
=  {\includegraphics[scale=0.5,valign=c]{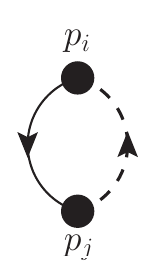}}~,
\end{equation}
to simplify the results of Feynman diagrams.

Since $\cAngle{ij}= \mathrm{sgn}(p_i^0p_j^0)\cSquare{ji}^*$,
we can also express the inner product in terms of the invariant mass and a phase
\begin{equation}
\cAngle{ij} \defequal e^{-i\varphi_{ij}} \sqrt{|s_{ij}|}~,
\qquad
\cSquare{ji} = \mathrm{sgn}(p_i^0p_j^0) e^{i\varphi_{ij}} \sqrt{|s_{ij}|} ~,
\qquad e^{-i\varphi_{ij}} =
\frac{p_i^\perp p_j^+ - p_j^\perp p_i^+}{\sqrt{|s_{ij}p_i^+p_j^+|}}~,
\label{eq:phase in pf and pb}
\end{equation}
which we used in \eqsref{eq:spinor phase eigenvectors},
(\ref{eq:spinor phase def}) and (\ref{eq:spinor phase def with alpha hel}).

\section{Tables with conventions and Feynman rules}
\label{sec:rosettastones_app}

We collect here the chirality-flow Feynman rules for the full (tree-level) Standard Model,
and compare them to other spinor-helicity notations.

\afterpage{
    \clearpage
    \thispagestyle{empty}
    \begin{landscape}
    \begin{table}
       \centering 
       \hspace*{-1.5cm}
       \begin{tabular}{c|c|c|c|c}
        \textbf{Species}  
        & \textbf{Index} & \textbf{Bra-ket} & \textbf{Feynman} & \textbf{Chirality-flow}\\ \hline 
        $u^+(p)$ 
        & 
        $\begin{pmatrix} -e^{-i\varphi}\sqrt{\al}\tla_q^\da \\ 
        \phantom{-e^{-i\varphi}\sqrt{\al}} 
        \la_{p^\flat,\al}\end{pmatrix}$&  
        $\begin{pmatrix} -e^{-i\varphi}\sqrt{\al}|q] \\ 
        \phantom{-e^{-i\varphi}\sqrt{\al}}
        |p^\flat \rangle \end{pmatrix}$ & 
        \rule{0cm}{0.8cm}
        \includegraphics[scale=0.4]{Jaxodraw/uPl_massive} &  $\begin{pmatrix}
-e^{\!-i\varphi}\!\sqrt{\alpha}
\raisebox{-5.5pt}{\includegraphics[scale=0.375]{./Jaxodraw/CR/ExtSpinorAntiDotted_q}}
\hspace{-1.75ex}
\\
\phantom{-e^{\!-i\varphi}\!\sqrt{\alpha}}
\raisebox{-2.5pt}{\includegraphics[scale=0.375]{./Jaxodraw/CR/ExtSpinorAntiSolid_pflat}}
\hspace{-1.75ex}
\end{pmatrix}$
\\
        $u^-(p)$ 
        & 
        $\begin{pmatrix} \phantom{e^{i\varphi}\sqrt{\al}}
        \tla_{p^\flat}^\da\\ 
        e^{i\varphi}\sqrt{\al}\la_{q,\al} \end{pmatrix}$ & 
        $\begin{pmatrix} \phantom{e^{i\varphi}\sqrt{\al}}
        |p^\flat]  \\ 
        e^{i\varphi}\sqrt{\al}|q\rangle\end{pmatrix}$ & 
        \rule{0cm}{0.8cm}
        \includegraphics[scale=0.4]{Jaxodraw/uMi_massive} &  $\begin{pmatrix}
\phantom{e^{i\varphi}\!\sqrt{\al}}
\raisebox{-5.5pt}{\includegraphics[scale=0.375]{./Jaxodraw/CR/ExtSpinorAntiDotted_pflat}}
\hspace{-1.25ex}
\\
e^{i\varphi}\!\sqrt{\alpha}
\raisebox{-5.5pt}{\includegraphics[scale=0.375]{./Jaxodraw/CR/ExtSpinorAntiSolid_q}}
\hspace{-1.25ex}
\end{pmatrix}$ \\
        $\bar{u}^+(p)$ 
        & 
        $\begin{pmatrix}
         \tla_{p^\flat,\da} \, , & 
         -e^{i\varphi}\sqrt{\al}\la_q^{\al}
\end{pmatrix}$  & 
        $\begin{pmatrix} [p^\flat|\, ,  & 
        -e^{i\varphi}\sqrt{\al}\langle q|\end{pmatrix}$ & 
        \rule{0cm}{0.8cm}
        \includegraphics[scale=0.4]{Jaxodraw/uBarPl_massive} &  $\bigg(
\!
\raisebox{-5.5pt}{\includegraphics[scale=0.375]{./Jaxodraw/CR/ExtSpinorDotted_pflat}}
\hspace{-2ex}
\;,\;
-e^{i\varphi}\!\sqrt{\alpha}
\raisebox{-5.5pt}{\includegraphics[scale=0.375]{./Jaxodraw/CR/ExtSpinorSolid_q}}
\hspace{-2ex}
\bigg)$
\\
        $\bar{u}^-(p)$ 
        & 
        $\begin{pmatrix}
          e^{-i\varphi}\sqrt{\al}\tla_{q,\da}\, , &
           \la_{p^\flat}^{\al}
\end{pmatrix}$ & 
        $\begin{pmatrix} e^{-i\varphi}\sqrt{\al}[q|\, ,  & 
        \langle p^\flat| \end{pmatrix}$  & 
        \rule{0cm}{0.8cm}
        \includegraphics[scale=0.4]{Jaxodraw/uBarMi_massive} &  $\bigg(
e^{\!-i\varphi}\!\sqrt{\alpha}
\raisebox{-5.5pt}{\includegraphics[scale=0.375]{./Jaxodraw/CR/ExtSpinorDotted_q}}
\hspace{-2ex}
\;,
\raisebox{-5.5pt}{\includegraphics[scale=0.375]{./Jaxodraw/CR/ExtSpinorSolid_pflat}}
\hspace{-1.5ex}
\bigg) $
\\
        $v^+(p)$ 
        & 
        $\begin{pmatrix} \phantom{-e^{i\varphi}\sqrt{\al}} 
        \tla_{p^\flat}^{\da} \\ 
        -e^{i\varphi}\sqrt{\al}\la_{q,\al} \end{pmatrix}$ & 
        $\begin{pmatrix} \phantom{-e^{i\varphi}\sqrt{\al}}
        |p^\flat ] \\ 
        -e^{i\varphi}\sqrt{\al}|q\rangle \end{pmatrix}$ & 
        \rule{0cm}{0.8cm}
        \includegraphics[scale=0.4]{Jaxodraw/vPl_massive} &  $\begin{pmatrix}
\phantom{\,-e^{i\varphi}\!\sqrt{\alpha}}
\raisebox{-5.5pt}{\includegraphics[scale=0.375]{./Jaxodraw/CR/ExtSpinorAntiDotted_pflat}}
\hspace{-1.25ex}
\\
\,-e^{i\varphi}\!\sqrt{\alpha}
\raisebox{-5.5pt}{\includegraphics[scale=0.375]{./Jaxodraw/CR/ExtSpinorAntiSolid_q}}
\hspace{-1.25ex}
\end{pmatrix}
$
\\
        $v^-(p)$ 
        & 
        $\begin{pmatrix} e^{-i\varphi}\sqrt{\al}\tla_{q}^{\da} \\ 
        \phantom{e^{-i\varphi}\sqrt{\al}}
        \la_{p^\flat,\al} \end{pmatrix}$  & 
        $\begin{pmatrix} e^{-i\varphi}\sqrt{\al}|q] \\ 
        \phantom{e^{-i\varphi}\sqrt{\al}}
        |p^\flat\rangle\end{pmatrix}$ & 
        \rule{0cm}{0.8cm}
        \includegraphics[scale=0.4]{Jaxodraw/vMi_massive} &  $  \begin{pmatrix}
e^{\!-i\varphi}\!\sqrt{\alpha}
\raisebox{-5.5pt}{\includegraphics[scale=0.375]{./Jaxodraw/CR/ExtSpinorAntiDotted_q}}
\hspace{-1.75ex}
\\
\phantom{e^{\!-i\varphi}\!\sqrt{\alpha}}
\raisebox{-5.5pt}{\includegraphics[scale=0.375]{./Jaxodraw/CR/ExtSpinorAntiSolid_pflat}}
\hspace{-1.75ex}
\end{pmatrix}
$
\\
        $\bar{v}^+(p)$ 
        & 
        $\begin{pmatrix}
         -e^{-i\varphi}\sqrt{\al}\tla_{q,\da}\, , & \la_{p^\flat}^{\al}
\end{pmatrix}$ & 
        $\begin{pmatrix} -e^{-i\varphi}\sqrt{\al} [q| & 
        \langle p^\flat| \end{pmatrix}$ & 
        \rule{0cm}{0.8cm}
        \includegraphics[scale=0.4]{Jaxodraw/vBarPl_massive} & 
        $ \bigg(
-e^{\!-i\varphi}\!\sqrt{\alpha}
\raisebox{-5.5pt}{\includegraphics[scale=0.375]{./Jaxodraw/CR/ExtSpinorDotted_q}}
\hspace{-2ex}
\;,\;
\raisebox{-5.5pt}{\includegraphics[scale=0.375]{./Jaxodraw/CR/ExtSpinorSolid_pflat}}
\hspace{-1.5ex}
\bigg) $
\\
        $\bar{v}^-(p)$ 
        & 
        $\begin{pmatrix}
         \tla_{p^\flat,\da}\, , & e^{i\varphi}\sqrt{\al}\la_q^{\al}
\end{pmatrix}$ & 
        $\begin{pmatrix} [p^\flat| \, , & 
        e^{i\varphi}\sqrt{\al}\langle q|\end{pmatrix}$ & 
        \rule{0cm}{0.8cm}
        \includegraphics[scale=0.4]{Jaxodraw/vBarMi_massive} &  
        $\bigg(
\!
\raisebox{-5.5pt}{\includegraphics[scale=0.375]{./Jaxodraw/CR/ExtSpinorDotted_pflat}}
\hspace{-2ex}
\;,\;
e^{i\varphi}\!\sqrt{\alpha}
\raisebox{-5.5pt}{\includegraphics[scale=0.375]{./Jaxodraw/CR/ExtSpinorSolid_q}}
\hspace{-2ex}
\bigg)$
    \end{tabular}
   \captionof{table}{
    ``Rosetta stone'' 
    showing massive fixed-spin-axis fermions in several common notations. 
    To translate the Dirac spinors to the helicity basis, 
    replace $p^\flat \rightarrow p_\plus$,
    $q\rightarrow p_\minus$, and remove $\sqrt{\al}$.
   The phase $\varphi$ is defined in 
   \eqref{eq:spinor phase def with alpha}
   in the general spin basis 
   (see \secref{sec:diracspinors}) 
   or in \eqref{eq:spinor phase def with alpha hel}
   in the helicity basis 
   (see \secref{sec:helicity eigenstates}).
   Massless spinors are found by setting $\alpha=0$
   and replacing $p^\flat \rightarrow p$.
   } 
    \label{tab:Flow rules fixed-axis spinors}
    \end{table}
    \end{landscape}
    \clearpage
}

\afterpage{
    \clearpage
    \thispagestyle{empty}
    \begin{landscape}
    \begin{table}
        \centering 
        \hspace*{-2.5cm} 
        \begin{tabular}{c|c|c|c|c|c}
        \textbf{Vertex} & \textbf{Dirac} & \textbf{Index} & \textbf{Bra-ket} & \textbf{Feynman} & \textbf{Chirality-flow}\\ \hline
        \rule{0cm}{1.3cm}
        $gq\bar{q}$ & 
        $i\frac{g_s}{\sqrt{2}}t^a_{i\jbar}\gamma^{\mu}$ & 
        $ig_st^a_{i\jbar} \begin{pmatrix}
         0 &  \tau^{\mu,\da\be} \\
         \taubar^{\mu}_{\al\db} & 0
         \end{pmatrix}$ &  
        $ig_st^a_{i\jbar} \begin{pmatrix}
         0 &  \tau^{\mu} \\
         \taubar^{\mu} & 0
         \end{pmatrix}$ & 
          \includegraphics[scale=0.35,valign=c]{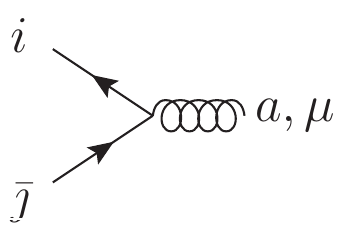}
         &
        $ig_st^a_{i\jbar} \begin{pmatrix}
         0& \raisebox{-0.35\height}{\includegraphics[scale=0.425]{./Jaxodraw/TauVertexHighb}}
         \\
         \raisebox{-0.4\height}{\includegraphics[scale=0.425]{./Jaxodraw/TauVertexLowa}} & 0
        \end{pmatrix}$ 
        \\ 
        \rule{0cm}{1.7cm}
        $ggg$ & 
        $V_3 \defequal i\frac{g_s}{\sqrt{2}} 
        (if^{a_1a_2a_3}) V_3^{\mu_1\mu_2\mu_3}$        
        & $V_3$ 
        & $V_3$ &
        \raisebox{-0.5\height}{ \includegraphics[scale=0.35]{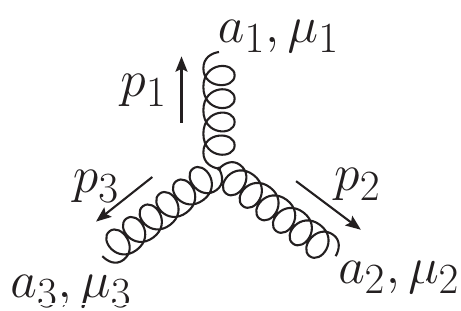}} & 
          $\begin{matrix} \hspace{-3.75cm}
          i\frac{g_s}{\sqrt{2}}(if^{a_1a_2a_3})\frac{1}{\sqrt{2}} \times \\
          \!\times \left( 
          \raisebox{-0.35\height}{ \includegraphics[scale=0.275]{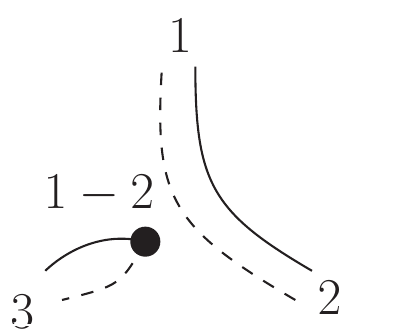}}
        \!\!\!+\!\!\!
        \raisebox{-0.45\height}{ \includegraphics[scale=0.275]{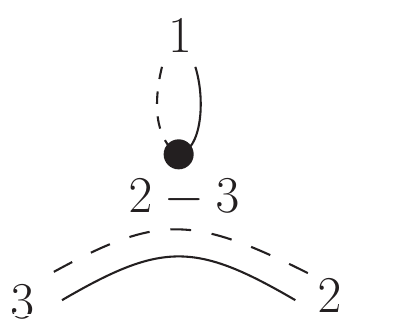}}
        \!\!\!+\!\!\!
        \raisebox{-0.4\height}{ \includegraphics[scale=0.275]{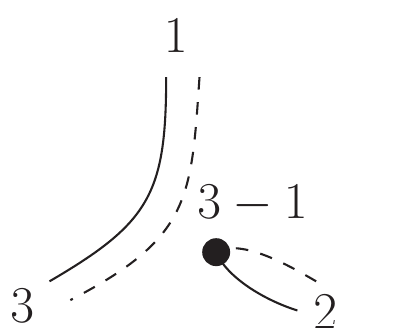}} \right)
          \end{matrix}           $  \\ 
    \rule{0cm}{1.7cm}    
    $gggg$ & 
        $\!\!\begin{matrix}
        \phantom{0} \\
        V_4\defequal\\
        \phantom{\left(\!\frac{g_s}{\sqrt{2}}\!\right)^{\!2}}
        \end{matrix}
        \begin{matrix}
        \phantom{0} \\
        i\!\left(\!\frac{g_s}{\sqrt{2}}\!\right)^{\!2}
        \!\!\!\!\!\sum\limits_{Z(2,3,4)}\!\!\!\!(if^{a_1a_2b})
        (if^{ba_3a_4}) \\
        \times(g^{\mu_1\mu_3}g^{\mu_2\mu_4}-
        g^{\mu_1\mu_4}g^{\mu_2\mu_3})
        \end{matrix}\!\! $ 
        & $V_4$  & $V_4$ &  
         \raisebox{-0.5\height}{ \includegraphics[scale=0.30]{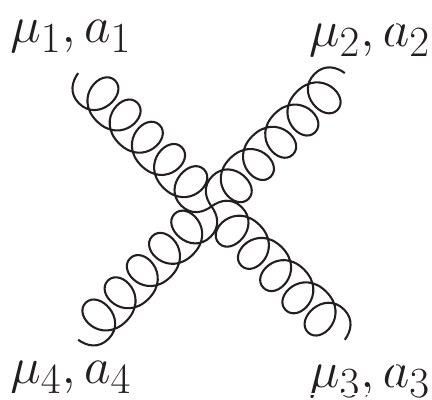}} &  
        $\begin{matrix} \hspace{-1.5cm}
         i\left(\!\frac{g_s}{\sqrt{2}}\right)^{\!2}\!\!\!\sum\limits_{Z(2,3,4)}\!\!
        (if^{a_1a_2b})(if^{ba_3a_4}) \times \\
        \times \left(\raisebox{-0.4\height}{ \includegraphics[scale=0.2]{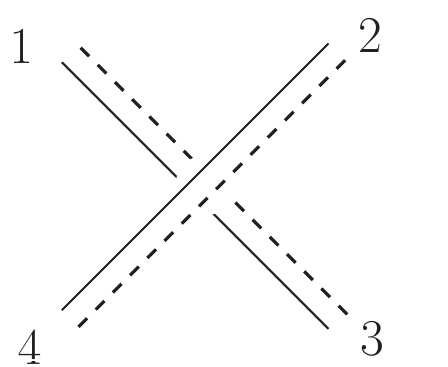}} 
        \!\!-\, \raisebox{-0.4\height}{ \includegraphics[scale=0.2]{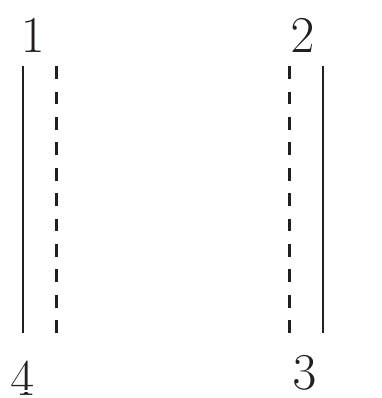}}\right)\!\!        
        \end{matrix} $ 
        \\ 
\rule{0cm}{1cm}        &  &  &  & 
        &  
        $ \begin{matrix} \hspace{-1.5cm} = i\left(\!\frac{g_s}{\sqrt{2}}\right)^{\!2}\!\!\!\sum\limits_{S(2,3,4)}\!\!\!\trace\big(t^{a_1}t^{a_2}t^{a_3}t^{a_4}\big) \times \\
       \times \left( 2 \!\!\raisebox{-0.4\height}{ \includegraphics[scale=0.2]{./Jaxodraw/4GluonVertex_Kin13}} 
       -\; \raisebox{-0.4\height}{ \includegraphics[scale=0.2]{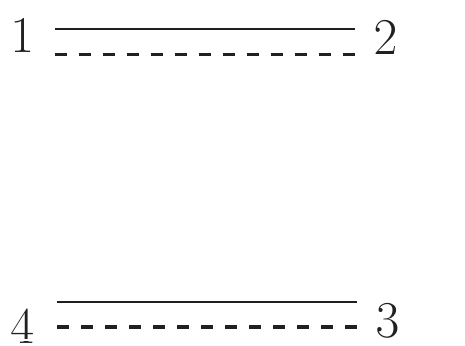}} 
       - \;\;\raisebox{-0.4\height}{ \includegraphics[scale=0.2]{./Jaxodraw/4GluonVertex_Kin14}} \right)
        \end{matrix}$  \\ \hline
        \textbf{Propagator} & \textbf{Dirac} & \textbf{Index} & \textbf{Bra-ket} & \textbf{Feynman} & \textbf{Chirality-flow}\\ \hline
        \rule{0cm}{1cm}        
        fermion & 
        $\frac{i\delta_{i\jbar}(p_\mu\ga^\mu + m)}{p^2-m^2}$
        & $\frac{i\delta_{i\jbar}}{p^2-m^2} 
           \begin{pmatrix} 
          m {\delta^{\da}}_{\db} & \sqrt{2}p^{\da \be} \\ 
         \sqrt{2}\bar{p}_{\al \db} &  m {\delta_{\al}}^{\be}
        \end{pmatrix}$ 
        & $\frac{i\delta_{i\jbar}}{p^2-m^2} 
         \begin{pmatrix} 
         m & \slashed{p} \\ 
          \bar{\slashed{p}} &  m 
        \end{pmatrix}$ 
        & \raisebox{-0.1\height}{\includegraphics[scale=0.4]{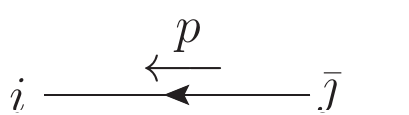}} 
        & $\frac{i\delta_{i\jbar}}{p^2-m^2} 
        \begin{pmatrix}
        m\includegraphics[scale=0.3]{./Jaxodraw/KroneckerDeltaDottedProp} 
        & \includegraphics[scale=0.3]{./Jaxodraw/FermionPropFlowg} 
        \\ 
        \includegraphics[scale=0.3]{./Jaxodraw/FermionPropFlowh} 
        & m\includegraphics[scale=0.3]{./Jaxodraw/KroneckerDeltaUndotted}
        \end{pmatrix}$ \\
        gluon & $-i\frac{\delta^{ab}g_{\mu\nu}}{p^2}$ & $-i\frac{\delta^{ab}g_{\mu\nu}}{p^2}$ & $-i\frac{\delta^{ab}g_{\mu\nu}}{p^2}$ & \hspace{-0.5ex}\raisebox{-0.2\height}{\includegraphics[scale=0.385]{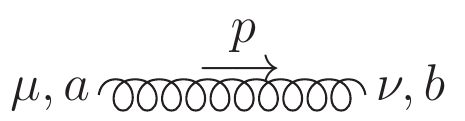}} & $ - \frac{i\delta^{ab}}{p^2}\raisebox{-0.25\height}{\includegraphics[scale=0.4]{./Jaxodraw/PhotonPropFlowa}}$ \ or \ $ - \frac{i\delta^{ab}}{p^2}\raisebox{-0.25\height}{\includegraphics[scale=0.4]{./Jaxodraw/PhotonPropFlowb}}$ 
    \end{tabular}
    \captionof{table}{The QCD ``Rosetta Stone" translating the chirality-flow notation to widely-used spinor-helicity notations. 
    Here $Z(2,3,4)$ denotes the set of cyclic permutations and
    $S(2,3,4)$ the set of all permutations of the integers $2,3,4$,
    and $V_3^{\mu_1\mu_2\mu_3} = 
      (p_1-p_2)^{\mu_3}g^{\mu_1\mu_2} 
    + (p_2-p_3)^{\mu_1}g^{\mu_2\mu_3}
    + (p_3-p_1)^{\mu_2}g^{\mu_3\mu_1}$.
    The external spinors and polarization vectors are trivially extended from 
    \tabsrefa{tab:Flow rules fixed-axis spinors}{tab:Flow rules comparison EW}
    respectively.
    For more information on the various forms of the four-gluon vertex,
    see section 5.1 in \cite{Lifson:2020pai}.
    The color factors are normalized such that
    $\trace\big(t^at^b\big) =\delta^{ab}$
    and 
    $if^{abc}= \trace\big(t^a[t^b,t^c]\big)$
    (see section 2 in \cite{Lifson:2020pai} 
    for details). 
  }
    \label{tab:Flow rules comparison QCD}
    \end{table}
    \end{landscape}
    \clearpage
}

\afterpage{
    \clearpage
    \thispagestyle{empty}
    \begin{landscape}
    \begin{table}
       \centering 
       \vspace*{-0.5cm}
       \hspace*{-3.60cm}
       \begin{tabular}{c|c|c|c|c|c}
        \textbf{Species}  & \textbf{Dirac} & \textbf{Index} & \textbf{Bra-ket} & \textbf{Feynman} & \textbf{Chirality-flow}\\ \hline 
        $\eps_{+}^\mu(p)$ & 
        $\epsilon_{+}^{\mu}(p)$ & 
        $\frac{\tla_{p^\flat}^{\da}\la_q^\be}{\la_q^{\ga} \la_{p^\flat,\ga}}$
         or  
        $\frac{\la_{q,\be}\tla_{p^\flat,\da}}{\la_q^{\ga} \la_{p^\flat,\ga}}$ & 
        $\frac{|p^\flat]\langle q|}{\langle qp^\flat\rangle}$
         or 
        $\frac{|q\rangle[p^\flat|}{\langle qp^\flat\rangle}$ &
        \raisebox{-0.2\height}{\includegraphics[scale=0.4]{Jaxodraw/PhotonExtPl}} & 
        $\;\;\frac{1}{\langle q p^\flat \rangle}$
        \raisebox{-0.2\height}{\includegraphics[scale=0.4]{Jaxodraw/ZExtFlowPlaRev}}
        \hspace{-2ex} or 
        $\;\;\frac{1}{\langle q p^\flat \rangle}$
         \rule{0cm}{0.8cm}
        \raisebox{-0.2\height}{\includegraphics[scale=0.4]{Jaxodraw/ZExtFlowPla}}
        \hspace{-2ex} \\  
                \rule{0cm}{0.9cm}
       $\eps_{-}^\mu(p)$ & 
       $\epsilon_{-}^{\mu}(p)$ & 
       $\frac{\tla_q^{\db}\la_{p^\flat}^{\al}}{\tla_{p^\flat,\dga} \tla_{q}^{\dga}}$ 
        or 
       $\frac{\la_{p^\flat,\al}\tla_{q,\db}}{\tla_{p^\flat,\dga} \tla_{q}^{\dga}}$ 
       & $\frac{|q]\langle p^\flat|}{[p^\flat q]}$
        or  
       $\frac{|p^\flat\rangle[q|}{[p^\flat q]}$ & 
       \raisebox{-0.2\height}{\includegraphics[scale=0.4]{Jaxodraw/PhotonExtMi}} & 
       $\,\;\frac{1}{[ p^\flat q ]}$
       \raisebox{-0.20\height}{\includegraphics[scale=0.4]{Jaxodraw/ZExtFlowMia}}
       \hspace{-2ex} 
       or $\,\;\;\frac{1}{[ p^\flat q ]}$
       \raisebox{-0.20\height}{\includegraphics[scale=0.4]{Jaxodraw/ZExtFlowMiaRev}}
       \hspace{-2ex} \\
        $\eps_{0}^\mu(p)$& 
        $\epsilon_{0}^{\mu}(p)$ & 
       $\frac{\tla_{p^\flat}^{\db}\la_{p^\flat}^{\al} - 
       \alpha \tla_{q}^{\db}\la_{q}^{\al}}{m\sqrt{2}}$ 
       & $\frac{|p^\flat]\langle p^\flat| - 
       \alpha |q]\langle q|}{m\sqrt{2}}$ & 
       \raisebox{-0.2\height}{\includegraphics[scale=0.4]{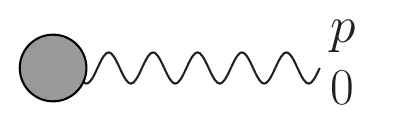}} & 
       $\,\;\frac{1}{m\sqrt{2}}$
       \raisebox{-0.20\height}{\includegraphics[scale=0.4]{Jaxodraw/ZExtFlow0dotRev}}\hspace{-2ex} 
       or $\,\;\;\frac{1}{m\sqrt{2}}$
       \raisebox{-0.20\height}{\includegraphics[scale=0.4]{Jaxodraw/ZExtFlow0dot}}\hspace{-2ex} \\

       \hline
        \textbf{Vertex} & \textbf{Dirac} & \textbf{Index} & \textbf{Bra-ket} & \textbf{Feynman} & \textbf{Chirality-flow}\\ \hline
        $Vff$ & $ie\gamma^{\mu}\left(C_LP_L + C_RP_R\right)$ & 
        $ie\sqrt{2} \begin{pmatrix}
         0 &  C_R\tau^{\mu,\da\be} \\
         C_L\taubar^{\mu}_{\al\db} & 0
         \end{pmatrix}$ &  
        $ie\sqrt{2} \begin{pmatrix}
         0 &  C_R\tau^{\mu} \\
         C_L\taubar^{\mu} & 0
         \end{pmatrix}$ & 
          \includegraphics[scale=0.35,valign=c]{Jaxodraw/fPhotonVertex}
         &
        \rule{0cm}{1cm}
        $ie\sqrt{2} \begin{pmatrix}
         0& C_R\raisebox{-0.35\height}{\includegraphics[scale=0.425]{./Jaxodraw/TauVertexHighb}}
         \\
         C_L\raisebox{-0.4\height}{\includegraphics[scale=0.425]{./Jaxodraw/TauVertexLowa}} & 0
        \end{pmatrix}$ \\
        $Sff$ & $ie\left(C_LP_L + C_RP_R\right)$ & 
        $ ie \begin{pmatrix}
        C_L\delta^{\db}_{~\da} & 0 \\
        0 & C_R\delta^{~\be}_{\al}
        \end{pmatrix}$ &
        $ie\begin{pmatrix}
        C_L & 0 \\
        0 & C_R
        \end{pmatrix}$ &
          \includegraphics[scale=0.35,valign=c]{Jaxodraw/vertFeynFFS}
         &
        \rule{0cm}{1cm}
        $ie \begin{pmatrix}
        C_L \raisebox{-0.4\height}{\includegraphics[scale=0.35]{Jaxodraw/vertChirFFS1_rosetta}} & 0 \\
        0 & C_R \raisebox{-0.4\height}{\includegraphics[scale=0.35]{Jaxodraw/vertChirFFS2_rosetta}}
        \end{pmatrix}
        $
        \\ \hline
        \textbf{Propagator}& \textbf{Dirac} & \textbf{Index} & \textbf{Bra-ket} & \textbf{Feynman} & \textbf{Chirality-flow}\\ \hline
        \rule{0cm}{1cm}        
        fermion & 
        $\frac{i(p_\mu\ga^\mu + m)}{p^2-m^2}$
        & $\frac{i}{p^2-m^2} 
           \begin{pmatrix} 
          m {\delta^{\da}}_{\db} & \sqrt{2}p^{\da \be} \\ 
         \sqrt{2}\bar{p}_{\al \db} &  m {\delta_{\al}}^{\be}
        \end{pmatrix}$ 
        & $\frac{i}{p^2-m^2} 
         \begin{pmatrix} 
         m & \slashed{p} \\ 
          \bar{\slashed{p}} &  m 
        \end{pmatrix}$ 
        & \raisebox{-0.1\height}{\includegraphics[scale=0.4]{Jaxodraw/FermionProp_Feyn}} 
        & $\frac{i}{p^2-m^2} 
        \begin{pmatrix}
        m\includegraphics[scale=0.3]{Jaxodraw/KroneckerDeltaDottedProp} 
        & \includegraphics[scale=0.3]{Jaxodraw/FermionPropFlowg} 
        \\ 
        \includegraphics[scale=0.3]{Jaxodraw/FermionPropFlowh} 
        & m\includegraphics[scale=0.3]{Jaxodraw/KroneckerDeltaUndotted}
        \end{pmatrix}$ \\
        vector  & $-i\frac{g_{\mu\nu}}{p^2-m^2}$ & $-i\frac{g_{\mu\nu}}{p^2-m^2}$ & $-i\frac{g_{\mu\nu}}{p^2-m^2}$ & \raisebox{-0.2\height}{\includegraphics[scale=0.385]{Jaxodraw/PhotonProp}} & $ - \frac{i}{p^2-m^2}\raisebox{-0.25\height}{\includegraphics[scale=0.4]{./Jaxodraw/PhotonPropFlowa}}$ \ or \ $ - \frac{i}{p^2-m^2}\raisebox{-0.25\height}{\includegraphics[scale=0.4]{./Jaxodraw/PhotonPropFlowb}}$ \\
        scalar & 
        $\frac{i}{p^2-m^2}$ & 
        $\frac{i}{p^2-m^2}$ & 
        $\frac{i}{p^2-m^2}$ & 
        \raisebox{-0.2\height}{\includegraphics[scale=0.385]{Jaxodraw/ScalarProp}} & 
        $\frac{i}{p^2-m^2}$ 
    \end{tabular}
       \captionof{table}{Part one of the electroweak ``Rosetta Stone'' translating
         the chirality-flow notation to widely-used spinor-helicity notations.
         To translate the polarization vectors to the helicity basis, 
         replace $p^\flat \rightarrow p_\plus$ and
         $q\rightarrow p_\minus$.
    Incoming polarization vectors of spin $\Js$
    are equal to outgoing ones of spin $-\Js$.
   Massless polarization vectors are given by replacing
   $p^\flat \rightarrow p$, and $q \rightarrow r $
   where $r$ is an unphysical reference momentum.
   Due to lack of space, 
   only the index and bra-ket forms of $\eps_0^\mu(p)$
   corresponding to the left chirality-flow diagram are given.
   The external spinors are given in \tabref{tab:Flow rules fixed-axis spinors}.
   Details about the polarization vectors, 
   vertices, and propagators are given in 
   \secsref{sec:polarizationvectors},
   \ref{sec:vertices}, and \ref{sec:propagators}
   respectively.
   } 
    \label{tab:Flow rules comparison EW}
    \end{table}
    \end{landscape}
    \clearpage
}

\afterpage{
    \clearpage
    \thispagestyle{empty}
    \begin{landscape}
    \begin{table}
        \centering 
        \hspace*{-2.5cm} 
        \begin{tabular}{c|c|c|c}
        \textbf{Vertex} & \textbf{Standard rule} & \textbf{Feynman} & \textbf{Chirality-flow}\\ \hline
        \rule{0cm}{1.0cm}
        \rule{0cm}{1.0cm}
        $VVS$ & 
        $ie C_{VVS}g_{\mu_1\mu_2}$        
        &
         \raisebox{-0.4\height}
         {\includegraphics[scale=0.35]{Jaxodraw/vertFeynVVS}} & 
          $ ieC_{VVS} \raisebox{-0.4\height}
          {\includegraphics[scale=0.35]{Jaxodraw/vertChirCurvedBoson}}$  \\ 
        \rule{0cm}{1.0cm}
        $SSS$ 
        & $ieC_{SSS}$ &
        \includegraphics[scale=0.35,valign=c]{Jaxodraw/vertFeynSSS} 
        & 
          $ ieC_{SSS} $  \\ 
          \rule{0cm}{1.0cm}    
        \rule{0cm}{1.0cm}
        $VVV$ & 
        $\begin{matrix}
        ieC_{V_1V_2V_3} \times \\
        \times \Big(
      g_{\mu_1\mu_2}(p_1-p_2)_{\mu_3} 
    + g_{\mu_2\mu_3}(p_2-p_3)_{\mu_1}
    + g_{\mu_3\mu_1}(p_3-p_1)_{\mu_2}\Big)
    \end{matrix}  $
        &
          \includegraphics[scale=0.35,valign=c]{Jaxodraw/vertFeynVVV}
           & 
          $ieC_{V_1V_2V_3}\frac{1}{\sqrt{2}}\left( 
          \raisebox{-0.35\height}{ \includegraphics[scale=0.275]{./Jaxodraw/3GluonVertex_HelFlow1}}
        \!\!\!+\!\!\!
        \raisebox{-0.45\height}{ \includegraphics[scale=0.275]{./Jaxodraw/3GluonVertex_HelFlow2}}
        \!\!\!+\!\!\!
        \raisebox{-0.4\height}{ \includegraphics[scale=0.275]{./Jaxodraw/3GluonVertex_HelFlow3}} \right)      $  \\ 
    \rule{0cm}{1.0cm}    
        \rule{0cm}{1.0cm}
        $SSV$ &  
        $ieC_{SSV}(p_1 - p_2)_\mu$ 
        &
        \raisebox{-0.4\height}{\includegraphics[scale=0.35]{Jaxodraw/vertFeynSSV}} & 
          $ ieC_{SSV} \frac{1}{\sqrt{2}}
  \raisebox{-0.4\height}{\includegraphics[scale=0.35]{Jaxodraw/vertChirSSV}} $  \\ 
    $VVVV$ 
        & $ie^2C_{V_1V_2V_3V_4}
        \Big(2g_{\mu_1 \mu_3} g_{\mu_4 \mu_2} 
    - g_{\mu_1 \mu_2} g_{\mu_3 \mu_4} 
    - g_{\mu_1 \mu_4} g_{\mu_2 \mu_3}\Big)$
        &  
         \includegraphics[scale=0.35,valign=c]{Jaxodraw/vertFeynVVVV} &  
         $ ie^2C_{V_1V_2V_3V_4} \left( 
       2\includegraphics[scale=0.35,valign=c]{Jaxodraw/vertChirVVVV1}
       -\includegraphics[scale=0.35,valign=c]{Jaxodraw/vertChirVVVV2}
       -\includegraphics[scale=0.35,valign=c]{Jaxodraw/vertChirVVVV3}
       \right) $ 
        \\ 
        \rule{0cm}{1.0cm}
        $VVSS$ & 
        $ie^2C_{VVSS} \,g_{\mu_1 \mu_2}$        
        &
        \includegraphics[scale=0.35,valign=c]{Jaxodraw/vertFeynVVSS}
        & $ ie^2C_{VVSS} 
        \includegraphics[scale=0.35,valign=c]{Jaxodraw/vertChirCurvedBoson} $  \\ 
        \rule{0cm}{1.7cm}
        $SSSS$ & 
        $ie^2C_{SSSS}$        
        &
        \includegraphics[scale=0.35,valign=c]{Jaxodraw/vertFeynSSSS} 
        & $ ie^2C_{SSSS} $  \\ 
    \end{tabular}
        \captionof{table}{Part two of the electroweak ``Rosetta Stone", translating
          boson vertices from the chirality-flow notation to the standard notation. 
    The coupling constants for every Standard Model process are found in \tabref{tab:all couplings}.
    The external spinors and polarization vectors are given in
    \tabsrefa{tab:Flow rules fixed-axis spinors}{tab:Flow rules comparison EW} respectively. 
  }
    \label{tab:Flow rules comparison EW bosons}
    \end{table}
    \end{landscape}
    \clearpage
}

\newpage
\begin{table}[h]
    \begin{subtable}[t]{0.45\textwidth}
    \begin{tabular}[t]{c|c|c}
       \textbf{Vertex} & \textbf{$C_L$} & \textbf{$C_R$}
       \\ \hline
       \rule{0cm}{0.5cm}    
       $ff\gamma$ & $Q_f$ & $Q_f$ \\
       \rule{0cm}{0.7cm}    
       $ffW$ & 
       $\frac{1}{\sqrt{2}\sin{\theta_W}}$ 
       & $0$    \\
       \rule{0cm}{0.7cm}    
       $ffZ$ & 
       $\frac{T_3^f-Q_f \sin^2 \theta_W}{\cos \theta_W \sin \theta_W}$
       & $-\frac{Q_f \sin \theta_W}{\cos \theta_W }$ \\
       \rule{0cm}{0.7cm}    
       $ffh$ & $-\frac{m_f}{2\sin \theta_W m_W}$
       & $-\frac{m_f}{2\sin \theta_W m_W}$
       \end{tabular}
       \caption{Fermion-boson vertices. The $ffW$ vertex \\ assumes diagonal flavor matrices.}
       \label{tab:couplings LR}
    \end{subtable}
    \begin{subtable}[t]{0.45\textwidth}
      \begin{tabular}[t]{c|c}
       \textbf{Vertex} & \textbf{Coupling} 
       \\ \hline
       \rule{0cm}{0.7cm}    
       $C_{WWh}$ & $\frac{m_W}{\sin \theta_W}$  \\
       \rule{0cm}{0.7cm}    
       $C_{ZZh}$ & $\frac{m_Z}{\sin \theta_W \cos \theta_W}
       = \frac{m_W}{\sin \theta_W \cos^2 \theta_W}$
       \\
       \rule{0cm}{0.7cm}    
       $C_{hhh}$ & $-\frac{3m_h^2}{2\sin \theta_W m_W}$
       \\
       \rule{0cm}{0.7cm}    
       $C_{\gamma W^+W^-}$ & $-1$
       \\    
       \rule{0cm}{0.7cm}    
       $C_{Z W^+W^-}$ & 
       $-\frac{\cos \theta_W}{\sin \theta_W}$
       \\
       \rule{0cm}{0.7cm}    
       $C_{W^+W^-W^+W^-}$ & 
       $\frac{1}{\sin^2 \theta_W}$
       \\    
       \rule{0cm}{0.7cm}    
       $C_{W^+ZW^-Z}$ & 
       $-\frac{\cos^2 \theta_W}{\sin^2 \theta_W}$
       \\    
       \rule{0cm}{0.7cm}    
       $C_{W^+ZW^-\gamma}$ & 
       $-\frac{\cos \theta_W}{\sin \theta_W}$
       \\    
       \rule{0cm}{0.7cm}    
       $C_{W^+\gamma W^-\gamma }$ & $-1$
       \\    
       \rule{0cm}{0.7cm}    
       $C_{WWhh}$ & $\frac{1}{2\sin^2 \theta_W}$
       \\    
       \rule{0cm}{0.7cm}    
       $C_{ZZhh}$ & 
       $\frac{1}{2\sin^2 \theta_W\cos^2 \theta_W}$
       \\    
       \rule{0cm}{0.7cm}    
       $C_{hhhh}$ & 
       $-\frac{3m_h^2}{4\sin^2 \theta_W m_W^2}$
       \end{tabular}
       \caption{Three- and four-boson vertices}
       \label{tab:couplings four bosons}
    \end{subtable}
    \caption{All coupling factors for electroweak interaction vertices. $T_f^3$ is the eigenvalue of the third isospin operator, 
    $Q_f$ is the electric charge and 
    $\theta_W$ is the Weinberg angle.}
    \label{tab:all couplings}
\end{table}

\bibliographystyle{JHEP}  
\bibliography{chiralityflow_massive} 

\end{document}